\documentclass{emulateapj-rtx4}
\usepackage{subfigure}
\usepackage{rotating}

\slugcomment{Accepted for Publication in ApJ Sept 4 2013}

\shorttitle{Shock-enhanced Line Cooling In Stephan's Quintet}
\shortauthors{Appleton et al.}

\begin{document}

\title{Shock-enhanced C$^+$ Emission and the detection of H$_2$O from Stephan Quintet's Group-wide Shock using {\it Herschel}\thanks{{\it Herschel} is an ESA space observatory with science instruments provided by European-led Principal Investigator consortia and with important participation from NASA.}}

\author{P. N. Appleton\altaffilmark{1}\email{apple@ipac.caltech.edu}  P. Guillard\altaffilmark{2,3}, F. Boulanger\altaffilmark{3}, M.E. Cluver\altaffilmark{4}, P. Ogle\altaffilmark{5}, E. Falgarone\altaffilmark{6}, G. Pineau des For{\^e}ts\altaffilmark{3}, E. O'Sullivan\altaffilmark{7}, P.-A. Duc\altaffilmark{8},  S. Gallagher\altaffilmark{9}, Y. Gao\altaffilmark{10}, T. Jarrett\altaffilmark{11}, I. Konstantopoulos\altaffilmark{4}, U. Lisenfeld\altaffilmark{12}, S. Lord\altaffilmark{1}, N. Lu\altaffilmark{1}, B. W. Peterson\altaffilmark{13},  C. Struck\altaffilmark{14}, E. Sturm\altaffilmark{15}, R. Tuffs\altaffilmark{16}, I. Valchanov\altaffilmark{17}, P. van der Werf\altaffilmark{18}, K. C. Xu\altaffilmark{1}}

\altaffiltext{1}{NASA Herschel Science Center, California Institute of Technology, Pasadena, CA 91125}
\altaffiltext{2}{Spitzer Science Center, California Institute of Technology, Pasadena, CA 91125}
\altaffiltext{3}{Institut d'Astrophysique Spatial\'e, Universite Paris Sud 11, Orsay, France}
\altaffiltext{4}{Australian Astronomical Observatory, Epping, NSW, Australia}
\altaffiltext{5}{NASA Extragalactic Database, IPAC, California Institute of Technology, Pasadena, CA 91125}
\altaffiltext{6}{Ecole Normale Sup\'erieure/Observatoire de Paris, France}
\altaffiltext{7}{Harvard-Smithsonian Center for Astrophysics, 60 Garden Street, Cambridge, MA 02138}
\altaffiltext{8}{Laboratoire AIM, Saclay, Paris, France}
\altaffiltext{9}{University of Western Ontario, Canada}
\altaffiltext{10}{Purple Mountain Observatory, Nanjing, China}
\altaffiltext{11}{Astronomy Department, University of Cape Town, Private Bag X3, , Rondebosch 7701, Republic of South Africa}
\altaffiltext{12}{Universidad de Granada, Granada, Spain}
\altaffiltext{13}{University of Wisconsin-Barron County, Rice Lake, WI 54868} 
\altaffiltext{14}{Iowa State University, Ames, IA 50011} 
\altaffiltext{15}{Max Planck Institute fur extraterrestische Physik, Munich, Germany}
\altaffiltext{16}{MPI-Kernphysik, Heidelberg, Germany}
\altaffiltext{17}{Herschel Science Center, ESAC, Madrid, Spain}
\altaffiltext{18}{Leiden Observatory, Leiden University, Leiden, Netherlands}

\begin{abstract}
 
We present the first Herschel spectroscopic detections of the
[OI]63$\mu$m and [CII]158$\mu$m fine-structure transitions, and a
single para-H$_2$O line from the 35 x 15 kpc$^2$ shocked  intergalactic
filament in Stephan's Quintet.  The
filament is believed to have
been formed when a high-speed intruder to the group
collided with clumpy intergroup gas.  Observations with the PACS spectrometer provide evidence
for broad ($>$ 1000 km~s$^{-1}$) luminous [CII] line profiles, as well as fainter [OI]63$\mu$m emission. SPIRE FTS
observations reveal water emission from the p-H$_2$O
(1$_{11}$-0$_{00}$) transition at several positions in the filament,
but no other molecular lines.  The H$_2$O line is narrow, and may be associated with denser intermediate-velocity gas experiencing the strongest shock-heating. The [CII]/PAH$_{tot}$
and [CII]/FIR ratios are too large to be explained by normal
photo-electric heating in PDRs. HII region excitation or X-ray/Cosmic Ray
heating can also be ruled out. The observations lead to the conclusion that a large fraction the molecular
gas is diffuse and warm.  We propose that the [CII], [OI] and warm
H$_2$ line emission is powered by a turbulent cascade in which
kinetic energy from the galaxy collision with the IGM is
dissipated to small scales and low-velocities, via shocks and
turbulent eddies. Low-velocity magnetic shocks can help explain both
the [CII]/[OI] ratio, and the relatively high [CII]/H$_2$ ratios
observed.  The discovery that [CII] emission can be enhanced,
in large-scale turbulent regions in collisional environments has implications for the interpretation of [CII] emission in
high-z galaxies.

\end{abstract}


\section{Introduction}

The Stephan's Quintet (hereafter SQ) compact galaxy group is unusual
among nearby compact groups because it contains a prominent ($\sim$35kpc
length) intergalactic filament, first discovered in the radio
continuum, but later found to emit optical emission lines and soft
X-rays consistent with a large-scale shock
\citep{sul01,xu03,trin05,osul09}.

\begin{figure}
\includegraphics[width=7.8cm]{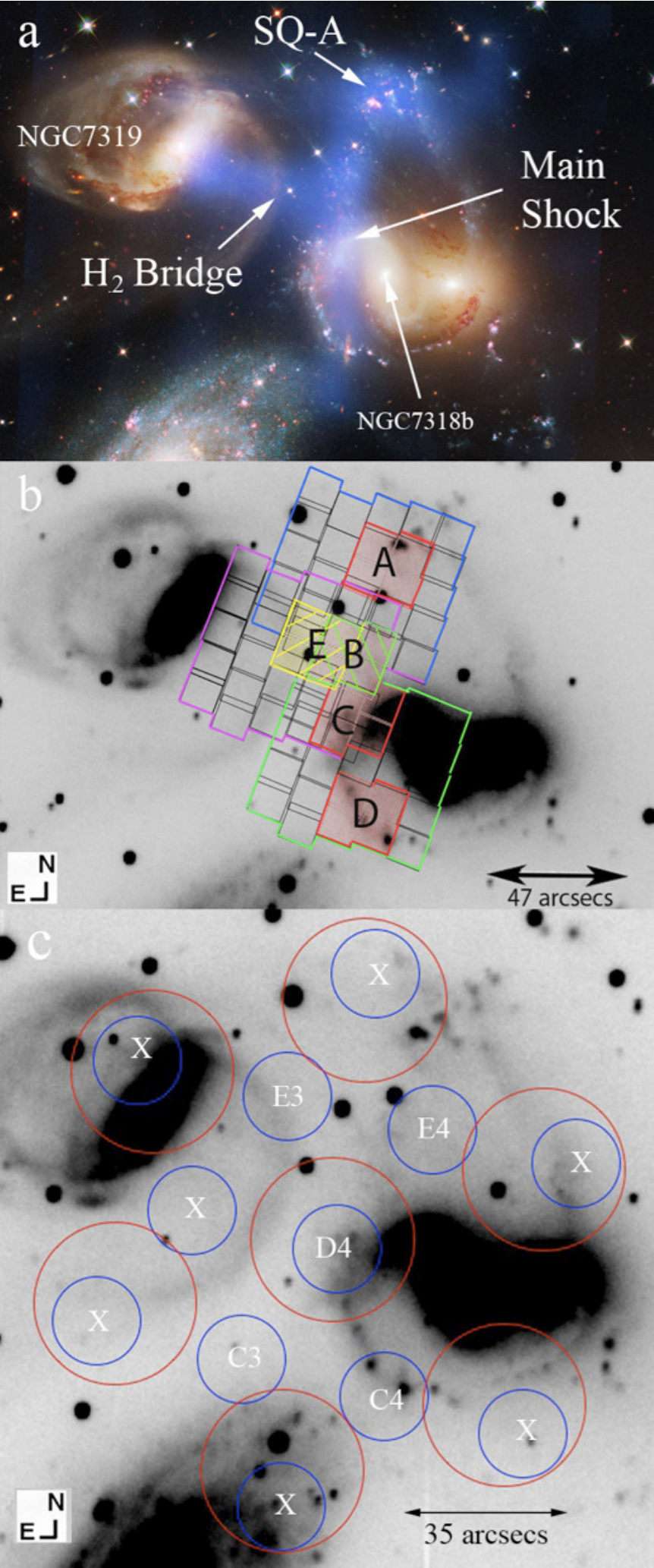}
\caption{\footnotesize{a) The distribution of warm molecular hydrogen (blue) superimposed on an
HSTWFC3 composite color image (red = H$\alpha$, green = V-band and
pale blue = B-band) of Stephan's Quintet. The H$_2$ outlines the
0-0S(1) distribution from \citet{m10}, b) R-band image of
Stephan's Quintet showing the outline of the three pointings of the 5
x 5 spaxel PACS IFU (blue, green and magenta). The lettered regions (A
through E) show the extraction boxes discussed in the text. Regions B
and E overlap, and were chosen to minimize contamination of the
shocked regions by faint star forming emission, c) SPIRE FTS
footprints superimposed on the R-band image. Blue and Red circles show the SSW
and SLW detector beam positions respectively. Those SSW detectors with
letters inside, denote those detectors in which the
pH$_2$O 1$_{11}$-0$_{00}$ molecule was detected (see Table 1 and 2)}. Those with a white 'X' show non-detections. 
SSWC3 was only marginally detected. Upper limits for
CO lines at position SLWC3 which is cospatial with SSWD4 is given in Table 2}

\label{footprint}
\end{figure}

The best explanation
for the formation of this giant structure is that it represents a
region of highly excited shocked gas caused by the collision of a
high-speed ``intruder'' galaxy, NGC 7318b, with a pre-existing tidal
filament generated in the past within the group \citep[see discussion by][] 
{mole98,sul01,will02}. Numerical models of
the past history of SQ support this picture 
\citep{rena10,sun12,gen12}.

The discovery of powerful, L(H$_2$) $>$ 10$^{42}$ ergs s$^{-1}$, broad
($\sim$800km s$^{-1}$ wide) pure-rotational mid-IR emission lines from warm
molecular hydrogen and [SiII]$\lambda$34.8$\mu$m \citep[see Fig.1a]{a6,m10}
provided the first glimpse of the tremendous
dissipation of energy in the shock--the H$_2$ lines significantly
out-shining the soft X-ray luminosity, and optical line emission. The
cooling time of the H$_2$ lines is so short that in-situ heating is
required to explain the emission. \citet{p9} provided an
explanation for the similar distribution of warm H$_2$ and X-ray
continuum in terms of the dissipation of energy caused by the
collision of the intruder with multi-phase intergroup gas. In this
model, X-rays are created in low-density regions which are shock-heated
to high temperatures, whereas the
H$_2$ emission arises in denser pockets of gas that survive the
passage of the main shock. Additional  evidence \citep{p12} for
a multi-phase shocked medium has come from the discovery of broad-line 
CO (1-0, 2-1 and 3-2) emission from the filament. Molecular gas is seen at 
velocities ranging from that of the ``intruder'' galaxy NGC7318b 
V$_{helio}$ = 5774 km s$^{-1}$) to that of the intergroup gas 
(V$_{helio}$ = 6600-6700 km/s). The
motions inferred from the CO kinematics support the idea
that a significant amount of  kinetic energy is still present in the
filament. Dissipation of this energy can easily provide a plausible 
source of in-situ heating of the warm H$_2$ emission. 

Given the potential importance of turbulence and shocks in dissipating
mechanical energy throughout the universe, a goal of the current
project is to quantify line cooling in an environment which is free of
the potentially confusing effects of star formation. If the picture of
a turbulent cascade of energy down from the galaxy-collision-scales to
the scale of small molecular-clouds is correct, energy may leak out at
different scales and densities. Our observations are aimed at
quantifying the importance of shocks and turbulence in the main far-IR ISM
cooling lines of [CII] and [OI]. The SQ filament is an obvious target
because we already have seen that molecular line emission is a large
fraction of the bolometric luminosity in the structure \citep{a6}. Furthermore,
Suzuki et al. (2011) suggested that [CII] emission might be contributing to a broad-band
160$\mu$m image obtained with AKARI. Indeed the approximate surface brightness levels inferred from their measurements for [CII] contamination are not far from the values we detect spectroscopically in this paper. 

In order to address some of these questions, we obtained
{\it Herschel} observations covering the wavelength range of important ISM
far-IR emission lines with PACS, as well as SPIRE observations which allow for
the potential detection of the higher-J transitions of CO (not
possible from the ground) which would probe denser and potentially warmer 
molecular clouds. 

Throughout this paper we adopt a distance to the main background group
(excluding the assumed foreground galaxy NGC 7320) of 94 Mpc \citet{Xu05}. 
At this distance, 10 arcsecs corresponds to a linear scale of
4.5 kpc.

\section{Observations and Data Reduction}
\begin{figure}
\includegraphics[width=7.5cm]{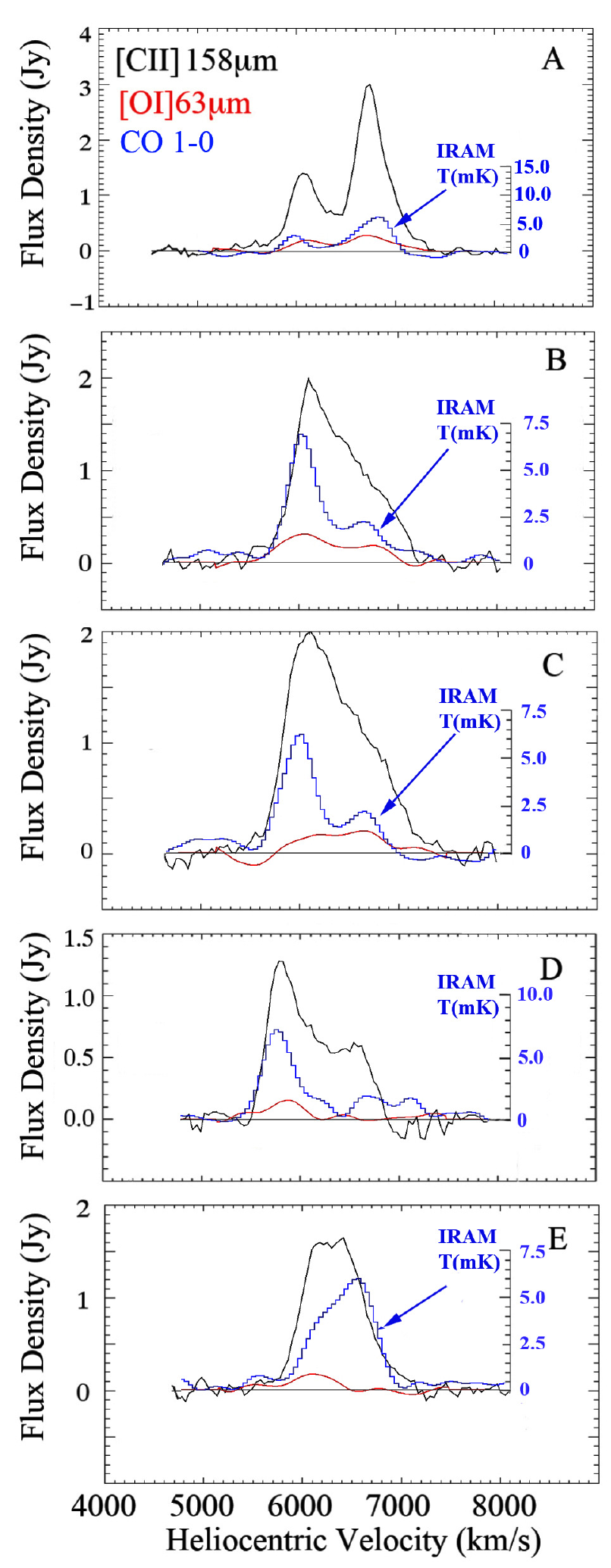}

\caption{\footnotesize{PACS spectra extracted from the five regions shown in Fig.1b. 
The
black line shows the spectrum of the [CII] line, and the red line
the [OI]63$\mu$m line. Each PACS spectrum represent the sum of the
emission over four native spaxels (final area 18.8 x 18.8 arcsecs$^2$). The blue line shows CO 1-0
spectra taken a positions close to the center of the PACS
extraction area with the IRAM 30-m telescope from
\citep{p12} with a circular beam size of 22 arcsec diameter. Although the beams are not perfectly matched (Region
B is offset by over half the FWHM of the IRAM beam-see text), it
is clear that the CO emission shares a lot in common with the
[CII] and [OI] emission.  The [OI] and CO data have been smoothed
to the same resolution of 235 km s$^{-1}$ as the [CII] line.}}
\end{figure}

Observations were made using the PACS integral field spectrometer
\citep{pog10} and the SPIRE Fourier Transform Spectrometer
(FTS; \citet{grif10}) onboard the Herschel Space Observatory
(Pilbratt et al. 2010) on 2011 Dec 7-8 and 2012 May 17 respectively,
as part of an open time program (PI Appleton\footnote{program name
  OT1$\_$pappleto$\_$1}). In addition, {\it Herschel} photometric observations 
from a companion paper (Guillard et al. 2013) will be used to provide 
far-IR continuum measurements in the current paper. 

For the PACS spectrometer, observations of the [CII]157.74$\mu$m and
[OI]63.18$\mu$m lines were made in the first and third-order gratings
using a short ``range-scan'' mode covering the redshifted wavelength
range 160.4-161.74 and 64.3-64.72 $\mu$m, with a velocity resolution
of $\sim$235 and $\sim$85 km s$^{-1}$ respectively. The first-order
and third order spectra are detected on independent red and blue
spectrometer arrays. The grating was stepped in high sampling mode
providing a heliocentric velocity coverage at the [CII] line of
4200-8500 km s$^{-1}$ and 5000-7700 km s$^{-1}$ for [OI] designed to
detect broad emission from the group.  The PACS Integral Field Unit (IFU) uses 
an image-slicer and reflective optics to project 5 x 5 spatial pixels 
(each 9.4'' x 9.4'' on the sky) through the spectrometer system over a total 
field of view of 47'' x 47''. Three separate ``pointed mode''
chop/nod observations were made (3 arcminute chopper throw) with 4 hrs
of integration time per pointing to cover the main parts of the SQ
filament (see Fig.1a,b). Two of the pointings cover the main north-south
structure of the molecular filament, and a third covers a connecting H$_2$ ``bridge'' 
between the main shocked filament and the galaxy NGC 7319 \citep[see ][]{m10}.

PACS data reduction was performed using the standard Herschel software
Herschel Interactive Processing Environment (HIPE)\footnote{
HIPE is a joint development by the Herschel Science Ground Segment Consortium, 
consisting of ESA, the NASA Herschel Science Center, and the HIFI, PACS and SPIRE consortia \citep{ott}.}
 user build 8.3, and
the results were later checked with HIPE 9.0 and 11.1 which have superior
flat-fielding capabilities (no significant differences were seen).
Data processing, included flagging and ignoring bad
pixels and saturated data, subtraction of chop ``on'' and ``off'' data,
division of the relative spectral response function, and the
application of a flat-field. The data were converted from standard
data frames to rebinned data cubes by binning these data in the
wavelength domain using default parameters (oversample = 2, upsample =
4) which samples the spectra at the Nyquist rate in the two bands.
Finally data for the two nods were averaged, resulting in a single
rebinned data cube for each of the three separate pointing positions
(Fig.1b).

As a check that the results were not affected by possible uncertainty
in the Relative Spectral Response Function (RSRF), and to check that the line fluxes and baselines were the
same using both methods (especially in the blue), we also ran these
data through a separate pipeline which normalizes the detector signals
to an average telescope background spectrum (e. g. see
\citet{gonz12} for brief description of method)-the
so-called flux normalization method.\footnote{At the time of writing,
the PACS Instrument Control Center recommends that this test be
performed as a check of the integrity of the final spectra. Although
the shape of the resulting spectra, could be affected by different
relative normalizations, the absolute calibration should be the same
for both methods.}  The results were essentially identical both in
baseline shape and flux density to less than a few $\%$. For both methods,  
the primary flux calibration is based on Neptune. Both methods should yield
absolute rms flux calibration uncertainties of 11$\%$ in both
bands\footnote{See PACS Spectrometer Calibration Document at
http://herschel.esac.esa.int/twiki/bin/view/Public/PacsCalibrationWeb}.

Because the PACS data were not taken in a fully
sampled manner, but rather as three separate pointed observations, we
cannot justify extractions of spectra from a combined map. Rather we
choose, more correctly, to extract individual spaxels from the
cubes. In order to obtain regions significantly larger than the FHWM
of the PSF at both wavelengths (9.3 arcsecs at 160$\mu$m and 3.7
arcsec at 64$\mu$m), we generously extracted 2 x 2 spaxel regions
(effectively areas 18.8 x 18.8 arcsec$^2$). These five large regions,
labled A through E, were selected for
spectral extraction. Since
the emission is observed to be quite smoothly distributed on the scale
of a few PACS spaxels, the extracted spectra should provide a
realistic measurement of the emission in the two lines.

The regions were selected to provide representative samples of the
main X-ray and H$_2$-defined shocked filament which runs nearly
North/South over a physical scale of 35 kpc: Region A through
D. Region E, was extracted in the direction of the feature which we
call the ``H$_2$ bridge'' (see Cluver et al. 2010), which previous
observations \citep{p12} have shown to contain broad CO lines indicating strong
turbulence. In order to ensure we could extract {\it Spitzer} IRS
spectra from the same regions as PACS, we could not extend our
extraction boxes too far to the north or south of the PACS IFU areas
without loosing coverage with the IRS Short-low module. The short-low
module provides crucial information about the strength of PAH
features, and proves to be important later in the paper. Region E was
chosen specifically to minimize the amount of possible star formation
activity in the bridge by inspection of both the 24$\mu$m and 11.3
$\mu$m PAH maps of Cluver et al. As a result, Region E has some
overlap with Region B).  The regions (A-E) were extracted from the
final rebinned cubes (slicedFinalCube product) of the level 2 data
using the cube analysis task in the Spectrum Explorer package in HIPE
9.1. 

SPIRE FTS observations were made in
the sparse-mapping single-pointing mode with 100 repetitions. This
resulted in 3.7 hrs on-source integration time. The SPIRE FTS has two
detector arrays, SSW and SLW, covering overlapping bands (194-313 $\mu$m and
303-671 $\mu$m respectively), and was used in the high-resolution mode. This provided a spectral resolution 370 $<$ R $<$ 1300 from the short to
longest wavelengths. The FTS data were processed using HIPE 9.0 user
reprocessing script with calibration product spire\_cal\_9\_0.  We retained
all the inner detectors up to the SSW/SLW co-aligned detector ring.
The data were processed to the final level where a point-source 
spectrum was derived per detector. 

Fig.1c shows the footprints of the FWHM of individual
detector-horn footprints superimposed on the R-band image of SQ. 

The Spitzer InfraRed Spectrometer (IRS) observations were made in
spectral-mapping mode and cover a large portion of the SQ
filament. Full details of these observations have been presented
elsewhere \citep{m10}. IRS data cubes constructed and
presented by Cluver et al. were used for comparison with the PACS IFU
data as follows. The Spitzer SL and LL data cubes, which had been
constructed at native spatial and spectral resolution using CUBISM
\citep{jd07}, were degraded at each wavelength within the cube
to an effective spatial resolution of 9.4 arcsecs designed to match
the resolution of the Herschel-observed [CII] line with PACS ($\lambda$
observed = 160$\mu$m) using a flux-conserving wavelength-dependant
convolution.  Spectra extracted from these convolved cubes formed the
basis for the comparison between the Spitzer and Herschel data
discussed below.

For the five selected PACS regions we also extracted photon count and
energy information from the deep (95 ks) {\it CHANDRA} X-ray map of
O'Sullivan et al. (2009).  The observations utilized the S3 CCD
detector of {\it Chandra} and the analysis followed the same
procedure as described in the above paper, except that the CIAO
package (version 4.5 with CALDB 4.5.5.1) was used to extract the PACS target regions. For each region
an absorbed Astrophysics Plasma Emission Code (APEC) model \citep{smi01} was fitted to the regions, providing best--fit values
for gas temperature, metallicity and X-ray luminosity. APEC calculates line and continuum emissivities for hot, optically thin plasma assumed to be in collisional ionized equilibrium, and draws on a library of over a million individual emission lines to build synthetic spectra which are used in the fit.

\section{Results}

\subsection{PACS Spectroscopy: Emission from [CII] and [OI]}

The PACS data cubes associated with the three
partially overlapping pointings revealed extended [CII] emission over most
of the region defined by the H$_2$ filament (e. g. the blue emission in
Fig.1a). Fig.2a-e show the extracted spectra from Region A-E (see Fig.1b) for
the [CII]158$\mu$m (solid black) and [OI]63$\mu$m (solid red)
displayed on the same radial velocity scale. To allow a better comparison, 
the [OI]63$\mu$m spectra were smoothed to the same velocity resolution as the
[CII] spectra (235 km s$^{-1}$). 

\begin{figure}
\includegraphics[width=8.5cm]{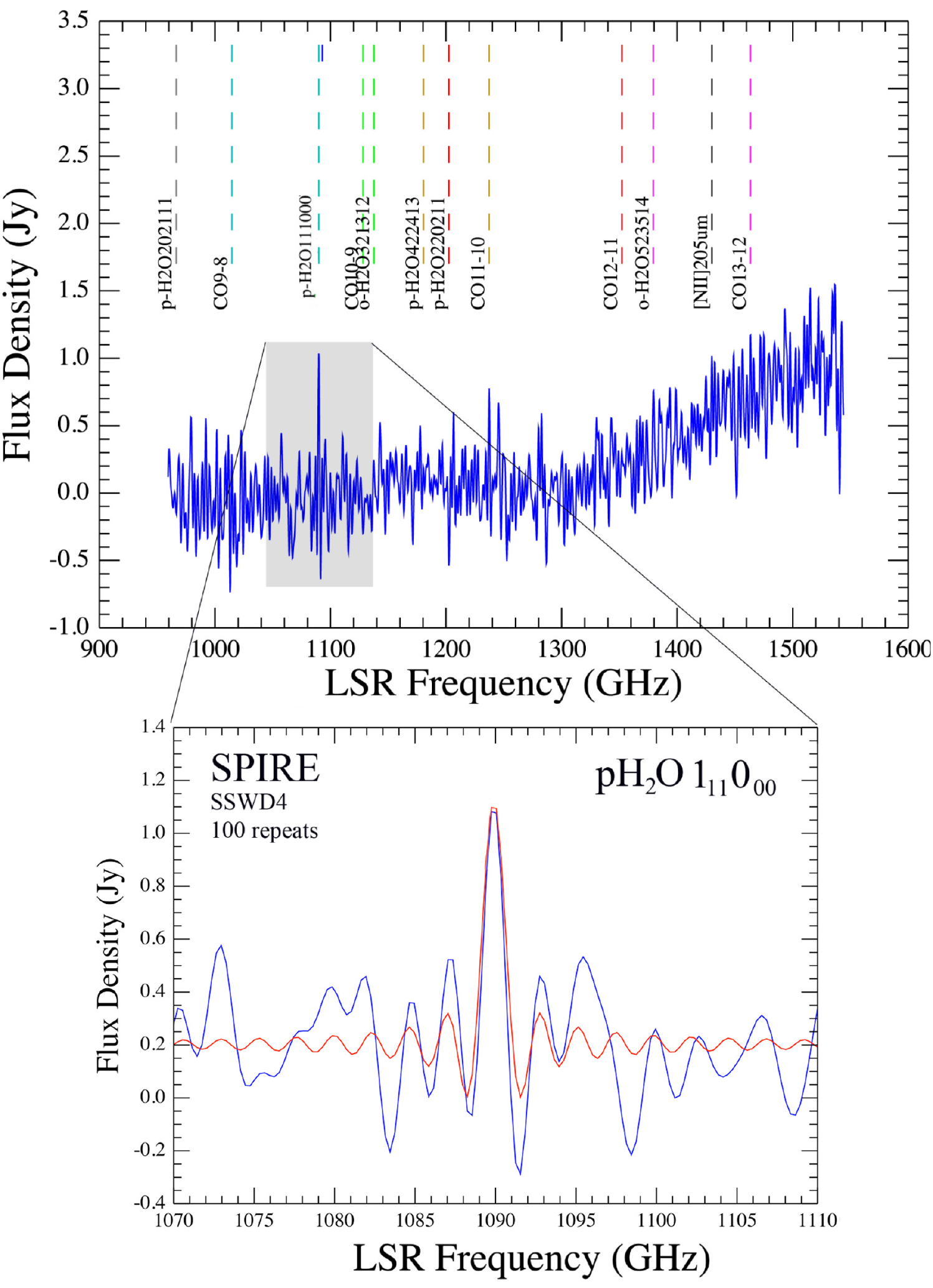}
\caption{\footnotesize{The SPIRE FTS spectrum at position SSWD4 (See Fig. 1c) covering 
the range 900-1600 GHz and the detection of the pH$_2$O line. Note the lack of 
detection of the [NII]205$\mu$m line at 1461 GHz. The zoomed inset shows a 
fit to the line with a SINC function (red line).}}
\end{figure}

\begin{figure}
\includegraphics[width=4.8cm]{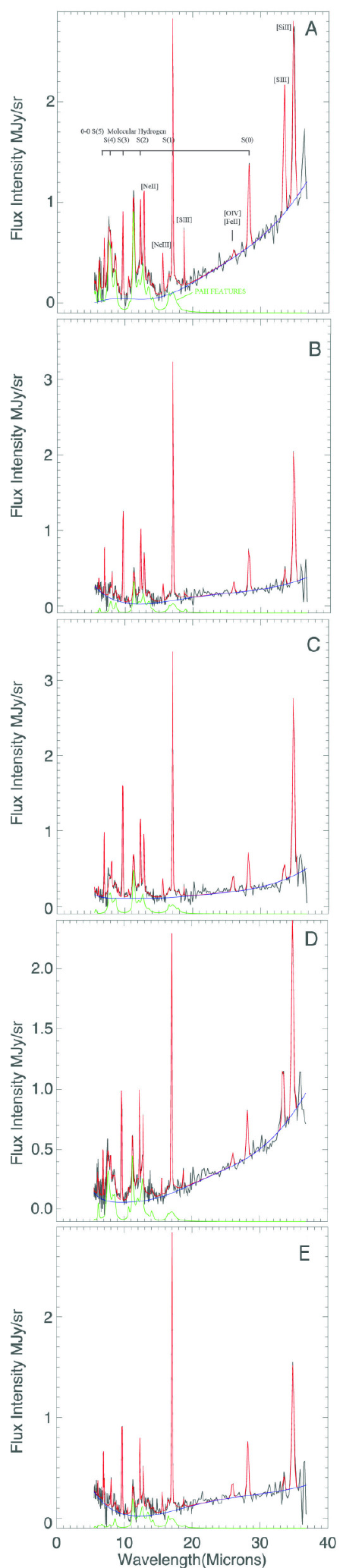}

\caption{\footnotesize{Spitzer IRS data taken from spectra cubes of the SQ
shock regions discussed in \citep{m10}. The five spectra
were extracted from the data cube over the same area as the
spaxels which contribute to the five spectra shown in Fig.2. Note
the very strong pure rotational H$_2$ lines, and weak PAH
features (shown in green based on the model fit from PAHFIT. The blue line represents the continuum fitted simultaneously with the lines and bands.}}
\end{figure}

The extracted spectra show several unusual features.  The [CII]
emission is strong, broad and asymmetric, with typical total line
widths exceeding 1000 km/s. This is consistent with broad
warm H$_2$ emission \citep{m10} observed with {\it Spitzer} with
considerably poorer spectral resolution. Fainter emission 
from [OI]63$\mu$m is also detected in Regions A, B and C,
covering the same velocity range as the [CII]. In D and E, the [OI]
seems to be only associated with the lower-velocity component of the
[CII] line. We note that the double-peaked profile evident in both the
[CII], CO and [OI] profiles for Region A is very similar to that seen
by \citet{will02} with the VLA in HI emission. HI was not detected in
the rest of the main north-south H$_2$ filament of Fig.1, i.e. sampled by  Regions B, C and D, nor in Region E. 

The spectra also resemble the single-dish observations of the CO
(1-0) transition of the cold molecular gas obtained with the IRAM 30m
telescope by \citet{p12}.  Except for Region B, where the IRAM beam is
offset from the PACS extraction center by about 14 arcsecs (more than
half the FWHM of the IRAM beam at 112.8 GHz-the observing frequency of
the CO line), the other pointings differ by no more than 4
arcsecs. These spectra are shown superimposed on the PACS spectra in Fig.2, and
have been smoothed to an effective spectral resolution of 235 km
s$^{-1}$ to match the [CII] resolution. Although the [CII] data do not
perfectly match the size and shape of the circular IRAM beam (FWHM
22''), the extracted square PACS spaxels in regions A, B and C show
similar spectral components to the CO. In Regions B and C, a bright
low-velocity component and a fainter high-velocity component are
evident, whereas for A, the situation is reversed. The [CII] emission
seems to fill-in the velocity space between the two main components of
the CO emission. The nearest IRAM spectrum to Region E is only offset
by 2.6 arcsecs in declination from the PACS spectrum and yet shows
some differences from the PACS extraction. Recent CO observations made
with PdB interferometer (Guillard et al. in preparation), show
significant velocity gradients across that region in the molecular
gas, and so the offset in the IRAM pointing may be responsible for the
different line shape between the [CII] and the CO. However, it is
clear that both the CO and the [CII] lines are broad
there\footnote{Region D in the smoothed IRAM spectrum contains a
higher velocity component not seen in the [CII] emission, but this
may be partly baseline uncertainty in the IRAM data.}. The overall
similarity between the [CII] line profiles, and the CO line profiles
in the main part of the filament suggests a kinematic connection
between the [CII] emission (and in some cases the [OI]63$\mu$m
emission), and the molecular gas.

In addition to being broad, the [CII] emission is stronger than the
[OI] emission (see Table 1 where the extracted line fluxes are
presented). 
We will argue later that this is consistent with a warm diffuse gas
heated by a network of low-velocity magnetic shocks (C-shocks), and/or
turbulence. The weakness of the [OI]63$\mu$m emission is not consistent with
strongly dissipative J-shocks \citep{drai83,holmc89,f12}.

Single and multiple Gaussian line fitting of the extracted [CII] (and
[OI] where appropriate) was performed using the ISAP package developed
for ISO \citep{eck98}.  Line fluxes for the decomposed Gaussian fits are given in 
Table 2. Regions B, C and D required 3 different velocity components
spanning the range from 6000 - 7000 km/s to provide a good fit to
these data--in most cases very large line widths were required.
Region C and E have the broadest fitted single components with FWHM
of 630 and 750 km s$^{-1}$ respectively, although the composite spectra span
over 1000 km s$^{-1}$ in total dispersion.  

\subsection{The SPIRE Spectra}

At each detector position, we fit simultaneously a polynomial
continuum and all the targeted lines with individual SINC profiles in
frequency. For each SINC profile, we initially fixed the FWHM at the
value for an unresolved line (1.44 GHz).  The targeted lines are the CO rotational transitions, known
water lines, [NII] 205 microns, [CI] 370 \& 609 microns.  The only
clearly detected line is the water line p-H$_2$O (1$_{11}$0$_{00}$) at
$\nu$$_{rest}$ 1113.3GHz (= 269.28$\mu$m),
which is detected S/N $>$5 in four SSW
detectors: SSWD4, SSWE3, SSWE4 and SSWC4. The SSW spectrum for
detector D4 is shown in Fig.3, where the zoomed-in region shows the
p-H$_2$O (1$_{11}$0$_{00}$) line, detected at a S/N of $\sim$10 (See Table 1). 
We refitted this line allowing the FWHM to
vary, but the line was still found to be unresolved at 324 km s$^{-1}$
(1.44 GHz) resolution.  This is in strong contrast to the [CII] and
[OI] emission which appears much broader.  The SPIRE fitted
heliocentric velocities and FWHM for the detected positions are provided 
in Table 2.  The heliocentric radial velocity of the H$_2$O line
for SSW D4 is 6457$\pm$120 km~s$^{-1}$, which places it
close to the line-center of the broad, but asymmetric [CII] emission from the same position (PACS Region C of Fig.2c).  Table 2
also provides the central velocity and FWHM for the other detections
in SSW E3, E4 and C4. Note that there appears to be a significant
radial velocity difference between the center of the SQ shock near SSW
D4 and the SSW C4 (shock south) and SSW E3 (shock-north) detector
positions. This follows the general trend of lower heliocentric
velocities in the southern part of the shocked filament (\citet{p12}).

Interestingly, no other H$_2$O lines are detected (Table 3 provides upper limits). 
On the para-H$_2$O ladder, the next highest ground-state transition is the 
p-H$_2$O (2$_{02}$ 1$_{11}$) line at 987 GHz, and on the otho-ladder, the 
o-H$_2$O (1$_{10}$ 1$_{01}$) at 557 GHz. Neither of these or other higher-order 
lines are detected at 3-sigma levels of  
$<$ 0.48-0.2 x 10$^{-17}$ W m$^{-2}$, 3 to 5 times lower than the detection of 
the pH$_2$O (1$_{11}$ 0$_{00}$) line. 

The [NII]$\lambda$205$\mu$m line ($\nu$$_{rest}$ =
1461.13GHz), a line often associated with HII regions, is not detected, 
nor are any of the higher-J CO lines, commonly found in SPIRE spectra
of higher-excitation galaxies detected
convincingly. Upper limits for these and other commonly encountered ISM lines
are also given for these lines in Table 3.

\subsection{Comparisons with the {\it Spitzer} IRS: Extraction of  H$_2$ and PAH Features}

{\it Spitzer} IRS observations obtained as part of a large spectral
mapping program (Cluver et al. 2010) contain important spatial
information about molecular hydrogen cooling and PAH emission across
the SQ filament.  For comparison with the PACS regions, we performed
matching extractions of the IRS data cubes covering
the same areas as the PACS regions A through E. The spectra, shown
in Fig.4, are strongly dominated by molecular hydrogen lines.

Fig.4 also shows the model result of running PAHFIT \citep{jdb07}
on the spectra. The red lines show the fits to atomic and molecular
lines (mainly H$_2$), and the green line under each spectrum shows the
fitted PAH features. The results of the fit, converted into line
fluxes are presented in Table 1.  It is clear from the spectra that
the H$_2$ lines, 0-0S(0) to S(5) dominate the spectrum of the SQ
filament, with faint emission from [NeII]12.8$\mu$m. Stronger emission
from the [SiII]34.8$\mu$m line is likely shock-excited (Cluver et
al. 2010).
 
The IRS spectra in Fig.4 show that in the filament, the 6.3 and 7.7$\mu$m PAH bands, usually associated with star formation are very weak, especially in PACS regions B, C and E. Faint 11.3$\mu$m  PAH is detected at most positions, although the
total power in all the PAH lines is small. In Table 1 we
tabulate the power in the 7-8$\mu$m PAH features (defined as the sum of the 7.4, 7.6, 7.8, 8.3, and 8.6 $\mu$m bands if present) , as well as the
integral over the main PAH bands (PAH$_{tot}$ defined here as the sum
of all the common PAH features from 6.3-17 $\mu$m).  We find ratios of
0.35 $<$ H$_{2tot}$/PAH$_{7-8}$ $<$ 5.2, and 0.19 $<$
H$_{2tot}$/PAH$_{tot}$ $<$ 1.2 , values that are incompatible with PDR
photoelectric heating efficiencies and observations of normal galaxies (see Section 5.3). This property, and the absence of a strong enough soft X-ray
flux led to the conclusion that the warm H$_2$ detected by Spitzer
must be shock excited \citep[see also][]{p9}. The values
with the lowest H$_2$/PAH ratios are Regions A and D, both of which
are known to contain some star formation. Here, some PDR heating of the
gas may be present, but not dominant.  

Unusually large H$_2$/PAH ratios are also found in the turbulent bridge region
between the Taffy galaxies \citep{brad12}, and in a sub-set of radio
galaxies where shock-excitation is believed to heat the H$_2$ to
several hundred K \citep{Og10,p12}, as well as in a sub-set of AGN-dominated
galaxies in the SINGS survey (Rousell et al. 2007).

Later in the paper we will 
compare the far-IR spectra
with the mid-IR spectral properties from these IRS extractions.

\subsection{Comparison with Far-IR Continuum Emission from {\it Spitzer} and {\it Herschel} imaging}

Observations of the IR continuum observations in SQ have been hampered
by poor spatial resolution at the longer wavelengths with ISO and
Spitzer \citep{xu03,m10,nat10,p10} New PACS and SPIRE observations
have recently been made \citep{p13}, and we borrow some of the results
from that paper to compare with the spectral line
properties. Photometry was extracted from archival Spitzer 24$\mu$m
\citep{m10}, and the new observations in the PACS 100 and 160$\mu$m
and SPIRE 250, 350 and 500$\mu$m bands. A full description of the
photometric Herschel observations is given in the companion paper of
\citet{p13}.

In order to gain the best possible measurement of the mid- to far-IR
luminosity from the PACS spectroscopic extraction regions, we followed
a two-step process. First we convolved the 24, 100, 160, 250 and
350$\mu$m maps to a spatial resolution appropriate for the 500$\mu$m
SPIRE image ($\sim$29 arcsecs FHWM). We then extracted an SED at these
five wavelengths from each of the regions A-E shown in Fig.1b.
This
allowed us to fit the 24-500$\mu$m SED with a modified black-body and a 
mid-infrared power-law (see Casey et al. 2012 for a description of the method). 
We computed the gas masses from the dust masses, assuming a dust-to-gas mass 
ratio of 0.006 (Xu et al. 2003). We call this the ``smoothed'' SED model.

\begin{figure}
\includegraphics[width=8.0cm]{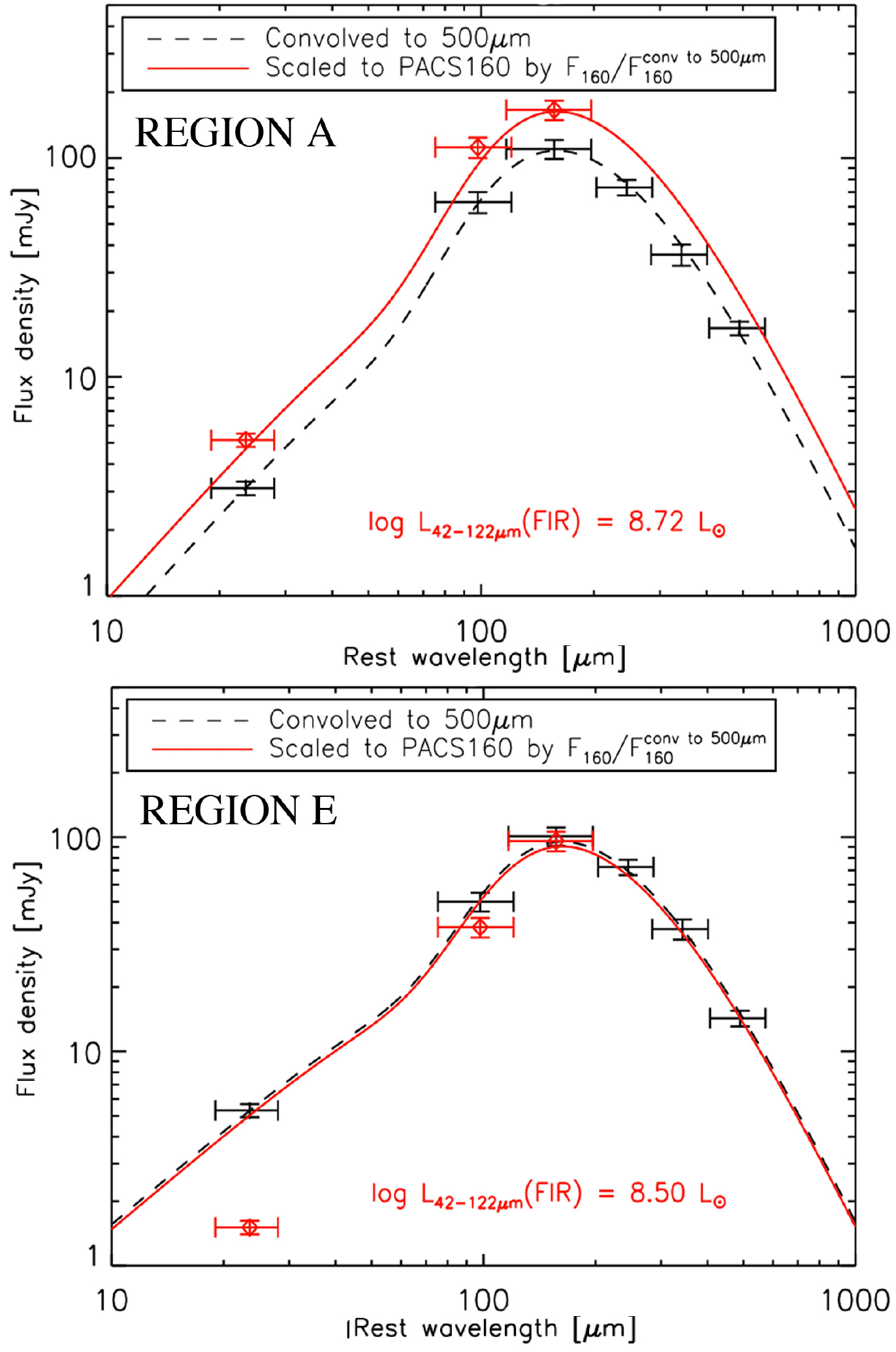}

\caption{\footnotesize{Example far-IR SEDs PACS Region A and E discussed in
more detail by Guillard et al. (2013). The black data
points represent extracted fluxes from 24, 100, 160, 250 and
350$\mu$m Spitzer, PACS and SPIRE images smoothed to the
resolution of the SPIRE 500$\mu$m map. The black dotted line
represents a fit (using the method of Casey 2012; See also Table 4)
to this is called ``smoothed model'' fit. The
red data points represents the measured fluxes  24, 100 and 160$\mu$m {\it Spitzer} and {\it Herschel}
PACS images (24 and 100$\mu$m smoothed to the resolution at 160$\mu$m). 
The red curve is the black-dotted line scaled to the flux at
full resolution PACS 160$\mu$m band to provide an estimate of FIR
luminosity at the same spatial resolution as the [CII] line (see text).For Region A, the flux increased in the full resolution SED because the emission is more point-like there.}}
\end{figure}

Secondly, in order to estimate the far-IR (FIR) fluxes
\citep{hel85,dale02} in each region, on an angular scale comparable
with the [CII] data, we re-extracted the 160 $\mu$m fluxes at the full
resolution of the 160$\mu$m map.  We then normalized the smoothed SED
model to the value at 160$\mu$m from the full-resolution map to derive
FIR flux per aperture over the range 42-122$\mu$m. Fig.5 shows the SED
obtained from the smoothed data, as well as the SED after tying it to
the full resolution 160$\mu$m data, for two examples Region A and
E. The effect of scaling the SED results in only a minor change when
the dust emission is extended (as in the case of Region E).  The low
flux value for the 24$\mu$m point in Region E when scaled to the
resolution of the 160$\mu$m observation (red point) is evidence for a
dearth of small grains (see later). The FIR properties measured in each
of the five PACS apertures calculated by this method is given in Table
4.

\begin{figure}
\includegraphics[width=8.5cm]{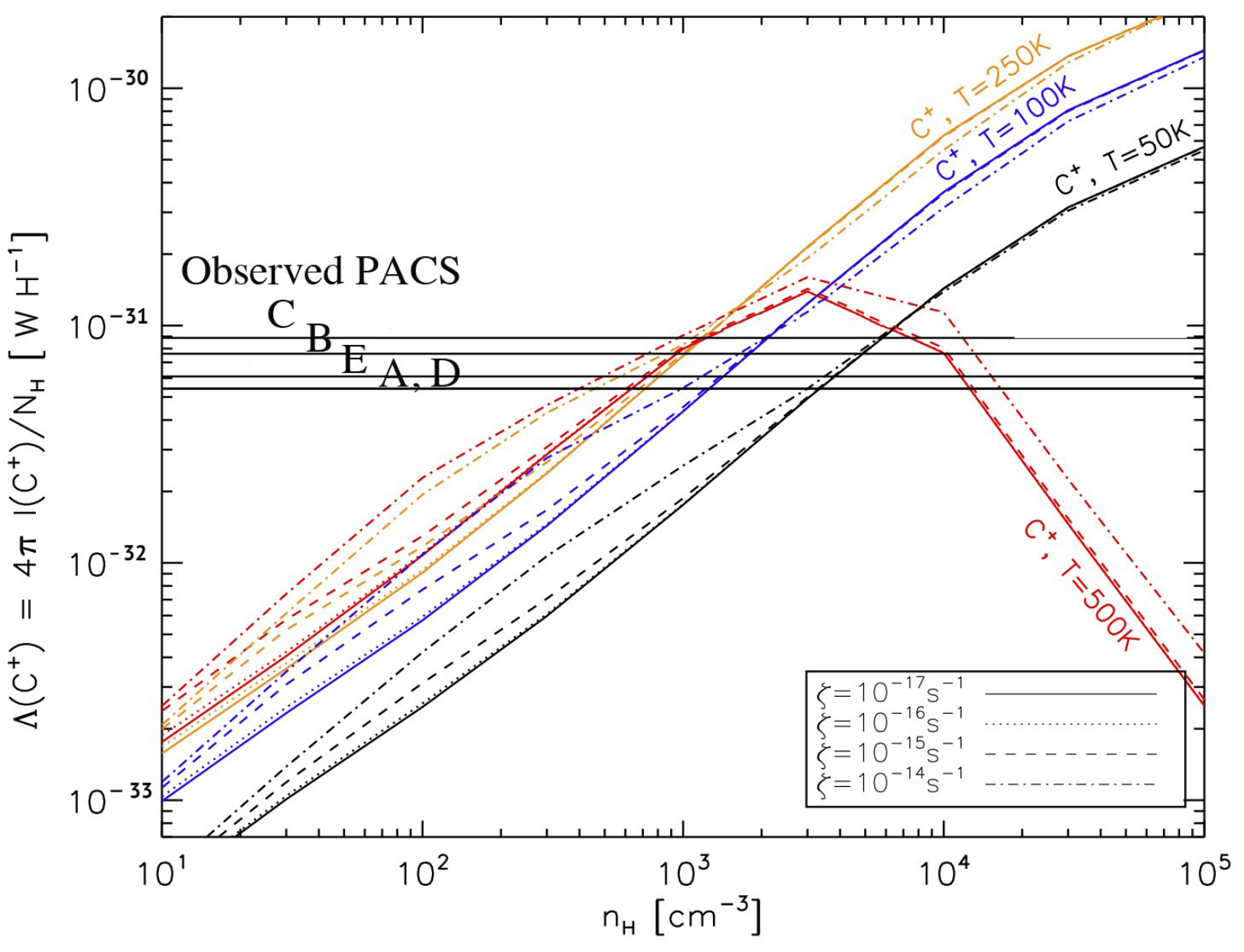}

\caption{\footnotesize{Theoretical models of the C$^+$ cooling rate as a function gas
density and temperature for various values of an assumed temperature and cosmic ray ionization rate (see text). Horizontal lines show the measured cooling rate of the C$^+$ observed by PACS for an assumed gas mass derived from the dust SED and an assumed gas to dust ratio of 0.006 (Xu et al. 2003).}} 
\end{figure}

\begin{figure}
\includegraphics[width=8.5cm]{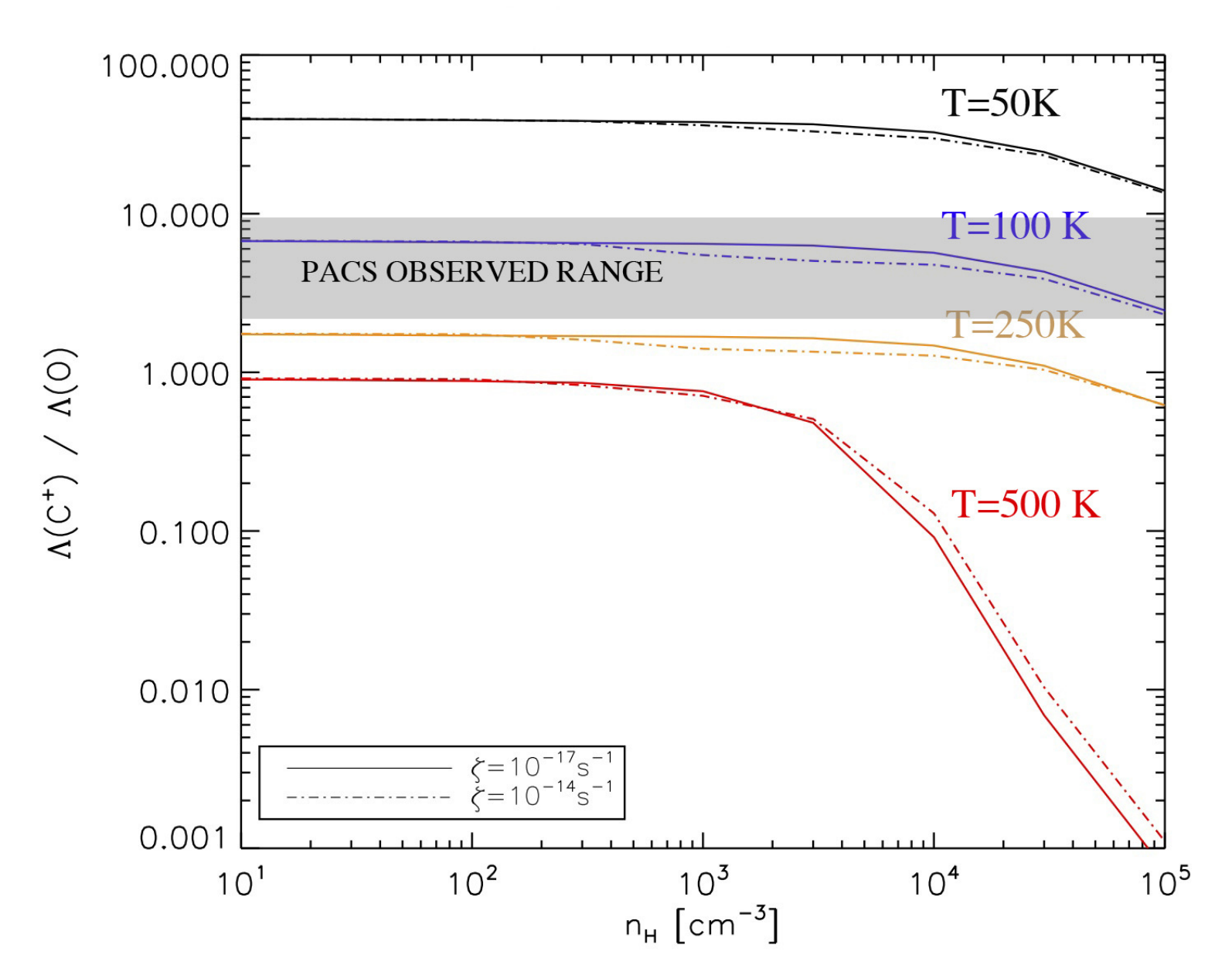}

\caption{\footnotesize{Models of the C$^+$/oxygen cooling rate ratio as a function gas
density and temperature for various values of the assumed cosmic ray ionization rate (see text). Horizontal grey bar denote the observed range in the ratio [CII]/[OI]63$\mu$m (Table 1) for each of the PACS extraction regions. The upper bound is Region E which has a large uncertainty. 
 The 63$\mu$m line dominates oxygen cooling.}} 
\end{figure}

Based on the FIR fluxes and [CII] line strength (summed over all the
velocity components detected within each PACS aperture) we calculate
the [CII]/FIR ratio, and this is given Table 1. It can be seen that
these ratios are unusually large, ranging from $\sim$0.04 (Region A \&
D) to 0.06-0.08 (Regions B, C and E). In general, [CII]/FIR ratios of
less than 1$\%$ are common in normal galaxies (see
Fig. 8). Interestingly, the regions that have the highest [CII]/FIR
ratios are also the same regions that show the most extreme [CII]/PAH
ratios. Those regions with known star formation (Region A and D) have
lower ratios, indicating that the star formation influences the
results by lowering both ratios.

We also measure approximately the intensity ratio of the [CII] to
CO (J=1-0) transition based on the spectra of Figure 2, given that
the extraction regions for the [CII] are square and not always
precisely aligned with the IRAM 30-m beam. The CO spectra were
converted to Jansky units from antenna temperature assuming a value
of 6.2 Jy/K, a value appropriate for compact emission smaller than
the 22 arcsec beam. We also scaled the [CII] extracted fluxes by a
factor of 1.54 to account for the difference in the areas of the [CII]
and CO extractions. If the emission is clumpy on the arcsecs scale,
this scaling may not be appropriate, and the [CII]/CO ratio would be lower.
The results range from 1000 $<$ I([CII])/I(CO J= 1-0) $<$ 1700 in Regions B through E,
and 2400 in Region A (which includes emission from the SQ-A star forming
region).  These values are on the low-end of the distribution of
values found in the Milky-Way and other nearby galaxies (e. g. Stacey
et al. 1991; 2010). One might be tempted to conclude that the low
values of I([CII])/I(CO) ratio point towards a low-excitation PDR
model, but this is not the case because the same regions also have
very elevated [CII]/FIR ratios placing them outside of any standard PDR
model (see Stacey et al. 2010; Fig. 3). Low metallicity is also unlikely to be the
explanation. The {\it average} metallicity of the gas (summed
over all the velocity components) is either mildly sub-solar or approximately solar (see Section 5.3 for a more complete discussion). Furthermore, our observations of the I([CII])/I(CO) ratio deviate significantly (to {\it lower} values) from those measured in the LMC/SMC. We will
show later that shock models are capable of reproducing the observed ratios. 

\section{Evidence for  Diffuse, Warm Molecular Gas from [CII] and [OI] Cooling Rates}
\begin{figure}
\includegraphics[width=8.5cm]{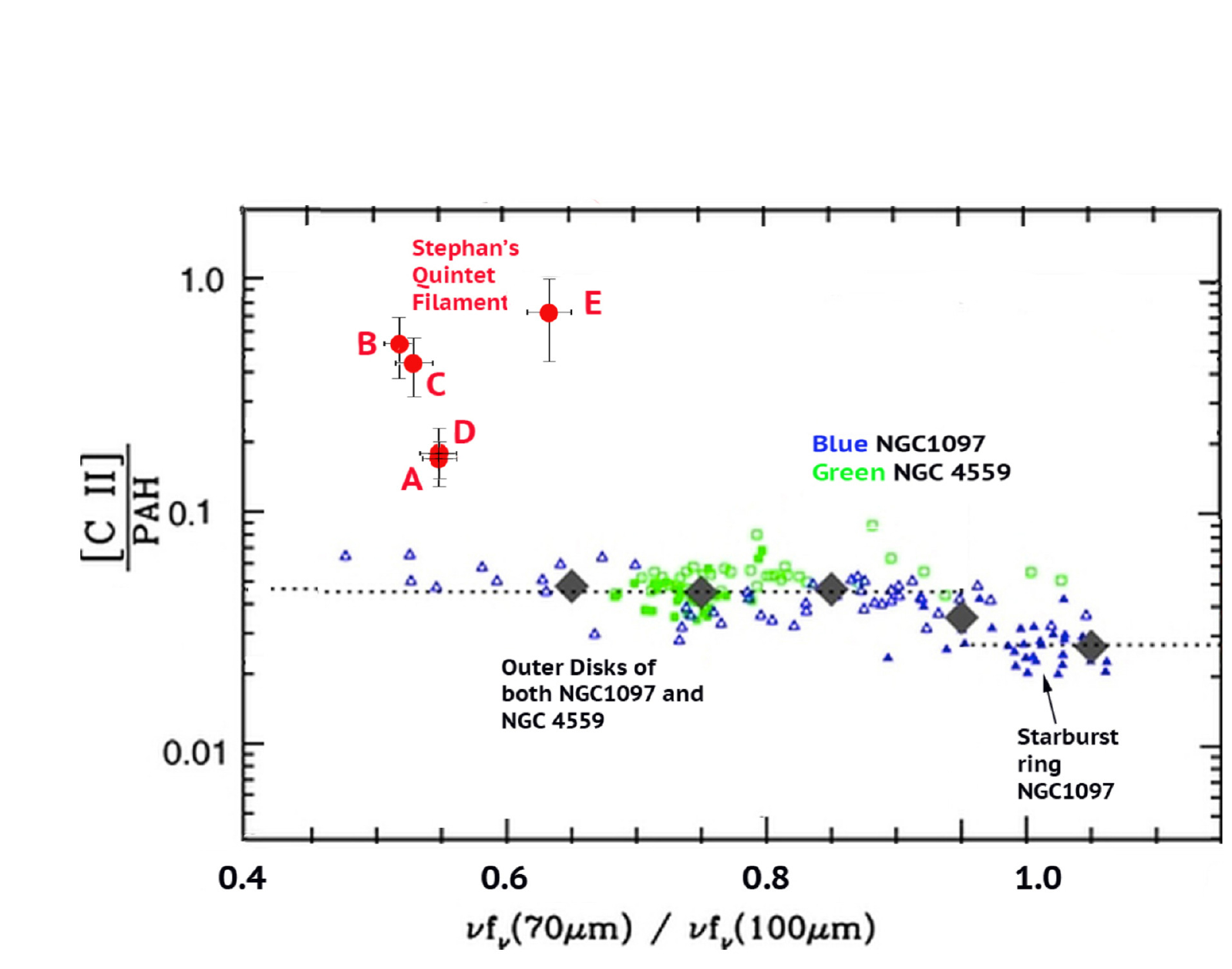}

\caption{\footnotesize{The [CII]$\mu$m/PAH$_{tot}$ ratio versus the FIR 70/100$\mu$m dust color for
NGC 1097 and NGC 4559 from \citet{crox12} KINGFISH program, with the SQ points from PACS
Regions A through E plotted. The photoelectric heating efficiency in
PDRs is dominated by PAH molecules, and [CII]/PAH ratio is unlikely to
exceed 3$\%$ for PAHs, and a smaller amount from small
and large grains according to \citep{hab04}. The SQ points are the values PAH$_{tot}$ as tabulated in Table 1, and the 70/100 $\mu$m colors are derived from the SED fits in Figure 5.
}}
\end{figure}

To constrain the gas temperature and density, we computed the
theoretical C$^+$ and oxygen cooling rates with a simple numerical model
to compare with our observations.  This code computes the gas
chemistry and thermal balance at a constant temperature and density,
assuming the carbon is all in the form of C$^+$. It makes no {\it a
  priori} assumptions about the source of heating which we will
consider in the next section. We considered 42 chemical species, and
use a chemical network of 285 reactions (a simplified version of the
one used in the shock model of \citet{flo03} and assume solar
metallicity. For the [CII] and total oxygen cooling rates we consider
the collisional effect of primarily H and H$_2$.  We also include
cosmic-ray heating, and parameterize its influence via the cosmic-ray
ionization rate.  The results of those calculations for several
different temperatures, densities and cosmic-ray ionization rates are
shown in Fig. 6 and 7. Fig.6 shows the [CII] cooling rate as a
function of gas density for four different temperatures and several
cosmic ray ionization rates. The figure shows that at n$_H$ $>$ 10$^3$
cm$^{-3}$, the [CII] cooling rate does not depend strongly on the
cosmic-ray ionization rate, because the gas rapidly recombines at
higher densities.  Above T$\sim$500K, the [CII] cooling drops at high
densities where oxygen cooling dominates. Horizontal lines denote the
observed line cooling in the five PACS regions A-E, as discussed
below.

Fig.7 shows the model ratio of the C$^+$ to total oxygen cooling (both the
[OI]63 and 145$\mu$m lines) as a similar function of density and
temperature. Although the model does not distinguish between the two
oxygen lines, [OI]63$\mu$m emission is the dominant coolant over the
range of densities and temperatures considered here.

\begin{figure}[b]
\includegraphics[width=8.0cm]{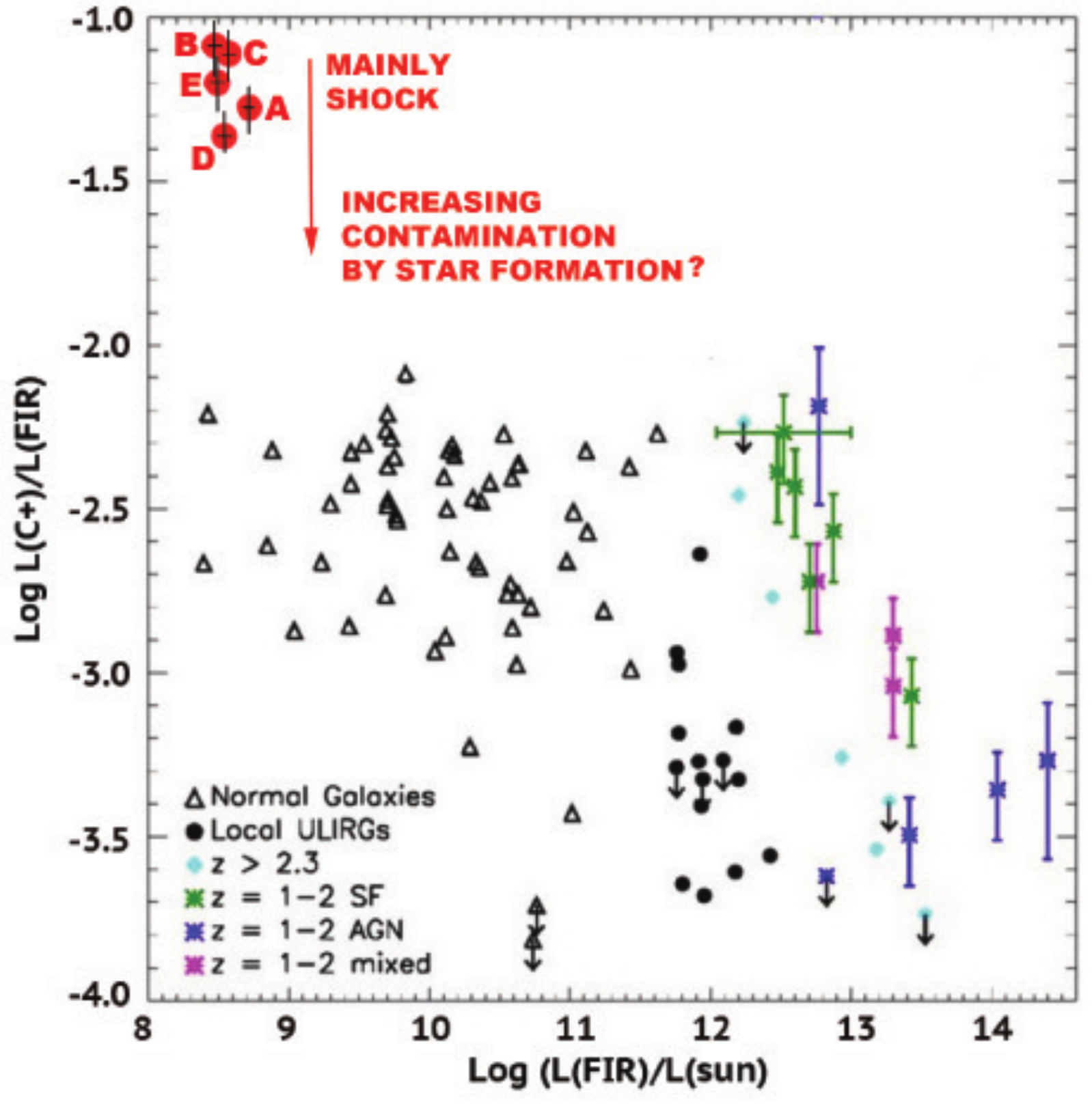}

\caption{\footnotesize{
A compilation by \citep{gord10} of the [CII]/FIR
luminosity versus the FIR luminosity for individual galaxies over
a wide range of luminosity and redshift. The SQ PACS regions A-E
are also plotted showing the extreme values for the [CII]/FIR
ratio, especially for the regions less contaminated by star
formation (regions B, C and E). Interestingly, those regions
exhibiting weak star formation fall closer to the distribution for
normal star forming galaxies. The arrow shows a highly schematic
mixing line as star formation becomes more dominant over shocks in
gas that is actively forming stars rather than in a highly
turbulent state.
}}

\end{figure}

How do these models compare with the observations of the [CII] lines?
The [CII] cooling rates for the 5 regions observed with PACS are
computed as $\Lambda (C^+) = L(CII) \times m_{\rm H} / M_{\rm gas}$,
where $m_{\rm H}$ is the Hydrogen mass and $M_{\rm gas}$ is the mass
of CII-emitting gas. Let us first assume that the total gas mass can
be derived from the far-IR observations (Guillard et al. 2013) based
on an assumed dust-to-gas ratio of 0.006 (e. g. from Xu et al. 2003).
These values are provided in Table 4 for each PACS region, and lie in
the narrow range 9.5 $\times$ 10$^7$ $<$ M$_{gas}$/M$_{\odot}$ $<$ 1.7
$\times$ 10$^8$, and a dust temperature of 22-24 K. If we assume
that all of this gas mass is involved in the observed [CII] cooling,
then the cooling rates, based on each PACS region will range from 5-9
$\times$ 10$^{-32}$ W H-atom$^{-1}$: the horizontal lines in Fig.6. 
The comparison of the modeled and observed cooling rates shows that
with this relatively high gas mass, the models and observation 
constrain the gas to be relatively warm (50 $<$ T $< $500 K), and of moderate density (500 $<$ n$_H$ $<$ 3-7 $\times$ 10$^3$).

The temperature can be further constrained if we also include the
observed [OI]63$\mu$m line cooling, and the ratio of the
[CII]/[OI]63$\mu$m obtained from Table 1 (ranging between values of 2 and
9) are shown on Fig.7. This indicates that the temperature of the gas
lies in a approximate range 90 $<$ T $<$ 200 K.  Combining the two diagnostics of
Fig.6 and 7, we can conclude that if all the gas inferred from the
dust SED takes part in the atomic line cooling, then the gas density
lies in a narrow range of densities around 10$^3$ cm$^{-3}$ and a
temperature of between 90 and 200 K.

It may be no coincidence that this
temperature is similar to the dominant temperature found by analysis of the
warm H$_2$ gas observed by Spitzer (see Cluver et al. 2010). 
For example, one could imagine that the warm H$_2$ mass
detected from {\it Spitzer} might be the dominant collisional heating
source for the [CII] emission. To consider this possibility, we derive 
the associated warm molecular hydrogen mass for each of the PACS 
extraction regions based on a simple two-temperature fit to the excitation diagrams of each region
derived from the {\it Spitzer} spectral extractions. 
These warm H$_2$ masses are presented in Table
5. These masses are only a factor of two smaller than
the gas masses estimated from the dust SED, 
ranging from 4-7 $\times$ 10$^7$ M$_{\odot}$ for PACS regions B,C and
E). \footnote{The  warm molecular hydrogen is an unusually
large fraction of the total gas mass. This conclusion was independently 
supported through the analysis of the CO emission (Guillard et al. 2012)}

Under the assumption that the C$^+$ gas is primarily collisionally heated
by the warm H$_2$, the cooling rate per H-atom (horizontal lines in
Fig.6) would double, leading to solutions with densities 1.5 $\times$
10$^3$ $<$ $\rho$ $<$ 4 $\times$ 10$^3$ cm$^{-3}$. The general
conclusion of the study of the cooling rates of the far-IR lines is
that the gas in the filament is both warm and diffuse.  It is thus
very plausible that the warm H$_2$ and the [CII] and [OI] emitting gas
are well mixed in approximate thermal balance.
 
\section{Possible Heating Sources for  the [CII]-emitting gas}

We have shown in the previous section that the [CII] emission 
is consistent with heating by a warm diffuse gas. What would be the source of heat for this gas?

\subsection{Extended distributions of HII Regions?}

Could the [CII] emission we observe be the result of collisional excitation
of the ground-state C+ ions by hot electrons in the plasma associated with a faint population of HII-region population scattered throughout the filament? Although there is evidence for a sparsely
scattered population of HII regions in the shock based on optical and
mid-IR measurements \citep{a6,m10,p10}, we can directly rule this
out from our own observations.  Firstly, our SPIRE observations,
which cover similar regions to those observed by PACS show no
detection of the [NII]205$\mu$m nebula line down to a 3-sigma
uncertainty of 2.7 x 10$^{-18}$ W m$^{-2}$ for detector SSW D4
(centered close to REGION C) and 2.0 x 10$^{-18}$ W m$^{-2}$ for the
average flux over the 4 inner detectors which were seen to show H$_2$O
emission. The [NII]205 line, because its ionization potential is 14.53 eV
\citep{ober06}, almost always arises in HII regions.  The lower limit
to the ratio of [CII] to [NII], after making a minor re-normalization
for the area of the SPIRE beams to that of PACS extraction area is,
I(C$^+$)/I([NII]205) $>$ ~36 for Region C. These values are significantly higher
than those expected for diffuse ionized emission over a realistic
range of densities by a factor of $>$ 7. 

This is not a surprise, since it has been known for some time that the
SQ filament contains only a scattering of compact HII regions along
its length \citep[see ][]{xu03,ober06,tra12}). These HII regions occupy a small
volume of the filament. Furthermore, the mass of gas required to
explain the strength of the [CII] line is far too large to be
associated with the observed compact HII regions. Guillard et
al. (2009) estimated the mass of ionized gas to be 1.2 $\times$
10$^{6}$ M$\odot$ over an aperture of 5.2 x 2.1 kpc$^2$, based on the emission measure of the H$\alpha$ line obtained across the filament by Xu et al. (2003), and assuming
reasonable conditions in the diffuse medium for the post-shocked
gas. This is at least an order of magnitude below the total warm molecular mass in the same aperture.

\subsection{X-ray heating?}

Given that the SQ filament is known to emit soft X-rays, it is
reasonable to ask whether X-rays could heat the gas in the
filament. We have shown in previous papers, based on older X-ray data,
that the X-ray emission is insufficient to heat the observed warm
H$_2$ emission over the area of the whole filament (Appleton et
al. 2006; Cluver et al. 2010). However, since then, new, much deeper {\it Chandra} data have
become available, which included a 97 ksec observations (O'Sullivan et
al. 2009) which have sufficient depth and spatial resolution to allow a direct comparison
with our [CII] and H$_2$ data extracted over our 18.8 x 18.8
arcsecs$^2$ PACS apertures. The results of the X-ray temperature and
abundance fits to the X-ray spectra allow us to obtain reliable X-ray
luminosities over these areas (see Table 6). From these results, we
can show that the [CII] to X-ray luminosity ratio,
L([CII])/L$_{Xsoft}$, varies from 3 to 11 depending on the region.
Thus the X-ray emission is far too faint to explain both the warm H$_2$
and the [CII] emission, since the efficiency for heating gas by X-rays is
a few percent at most. Only if the X-ray flux was underestimated significantly,
could X-rays be the main heat source. This is unlikely since fits to
the X-ray data do not suggest a large intervening column, nor does the
observed H$_2$ emission seen in the IRAM observation have enough
column density to provide significant X-ray obscuration.  Thus we can
be confident that the X-rays are not the primary heating source for
the main cooling lines. 

\subsection{Extended Photo-Dissociation-Regions (PDRs)?}

It has been known for many years that C$^+$ ions can be excited
in Photo Dissociation Regions (PDRs) in the diffuse ISM
\citep{tie85a,wat72,glas74,drai78,hol89,bak94}.  The dominant
mechanism is believed to be the heating of gas by photo-electric
ejection of electrons from PAH molecules and small grains.  The
warm gas (mainly the abundant HI and H$_2$) would then excite [CII]
emission collisionally.  Various authors have provided evidence that
PAHs dominate over both small and large grains in terms of
photoelectric heating efficiency \citep{wat72,holl99,hab04}.  The
heating of the diffuse gas may be further enhanced if the metallicity
of the gas is low, because UV photons can excite a larger volume at
smaller net G, thereby increasing the heating efficiency \citep[see
discussion of extended PDRs by][]{sue97,isr11}.

 One direct measure of the efficiency of photoelectric heating is the
 flux ratio of the [CII] to total emission from PAH features in the
 mid-IR spectrum [CII]/PAH$_{tot}$. Here we define PAH$_{tot}$ as the
 sum of the PAH feature fluxes from 6-17$\mu$m.  For diffuse regions
 in the Galaxy (e. g. Habart et al. (2003)), this is typically few
 percent, and has been directly measured recently in a variety of
 extragalactic environments \citep{ped12,crox12} ranging from
 2-5$\%$. Fig.8 (see also Table 1, column 7) shows the [CII]/PAH ratio
 for the five PACS regions plotted against far-IR color
 temperature. Also shown are spatially-resolved points from
 \citet{crox12} for NGC1097 and NGC 4559--two nearby galaxies observed
 in the Herschel KINGFISH program \citep{rob11}.  The plot emphasizes
 the unusually large ratios of [CII]/PAH$_{tot}$ in the SQ shock
 compared with normal diffuse emission in the outer parts of galaxies.

Our measured values of [CII]/PAH$_{tot}$ ratio range from values of
17-18$\%$ (Region A and D) to 40 to 70$\%$ in Regions B, C and E.  The
large [CII]/PAH ratios we observe, especially in Region E, strongly
argues against heating of the [CII] emission by photoelectron from PAH molecules in a
diffuse UV field as the primary mechanism for the [CII] emission,
since the efficiency of the photoelectric effect would have to be unusually large to explain the observed result.

\citet{m10} showed that the 24$\mu$m {\it Spitzer} emission from the
molecular shock is very weak, perhaps even absent at the shock-center
(near Region C) suggesting a dearth of small grains. This can now be
quantified more exactly with the new combined {\it Herschel} and 
{\it Spitzer}-derived SED \citep{p13}. The 24$\mu$m point for Region E
(scaled to the 160$\mu$m resolution--red-point in lower panel of
Fig.5) shows that it is extremely low, in sharp contrast to the same
point in Region A known to contain some star formation 
(upper panel of Fig.5).  Since small grains radiate
most efficiently in the 20-30$\mu$m range, it is unlikely that the
[CII] emission could arise from photoelectric heating from small
grains, since they seem depleted over much of the filament. Most of
the emission from Region E comes from long-wavelength emission,
presumably from larger grains where the photoelectric heating of [CII]
is the most inefficient (Bakes \& Tielens (1994).

PACS Region B, C and E also show unusually high [CII]/FIR ratios
(Table 1). These values ([CII] between 6-7$\%$ of the FIR) are much
larger than that seen in the diffuse ISM which is typically $<$ 1$\%$,
and exceeds by at least a factor of two the maximum theoretical
efficiency for photoelectric heating of [CII] by small grains
($\sim$3$\%$) discussed by \citet{bak94}. To put these values in a
more cosmic perspective we show in Fig. 9, the [CII]/FIR ratio versus the
FIR IR luminosity for the 5 extracted regions overlaid on a plot
shown by Stacey et al. (2010) for galaxies in both the nearby and
distant universe. Interestingly the Regions A and D, where the gas filament
is contaminated by star formation, show lower [CII]/FIR ratios,
whereas the other regions in SQ that lie in more ``pure shock''
environments away from known HII regions, have higher
ratios. This suggests that the signature of shocked or turbulently heated gas
can be masked by star formation--perhaps explaining why
such large ratios are not commonly seen in galaxies with powerful star formation. 

In Fig.10, we show one regime where the observations of the SQ shock
seems to fall closer to the norm seen in other galaxies. Here we plot
the [OI]63$\mu$m/[CII] ratio as a function of the FIR dust color
F(60$\mu$m)/F(100$\mu$m) from \citep{malh01} based on their work with
ISO for a set or normal galaxies. Although the points lie to the
extreme in color temperature (we determine these colors from the SED
fits of \citet{p13}), the points seem to form an
extension of those found for normal galaxies. Recent
spatially-resolved observations of M82 with PACS (Contursi et
al. 2013) show a tail of points which also extend into this
region. They mainly arise in the cooler outer parts of M82, with some
points possibly being associated with the starburst wind.  Malhotra et
al. interpreted their plot as an indication that [OI] is excited more,
with respect to C$^+$, as the dust temperature rises. Our results for SQ
appear to confirm that this trend remains true for the gas and dust in
the SQ filament, even though we have argued that process that excites
the [CII] (and presumably [OI]) is not UV radiation in a PDR. In some ways this is unfortunate, because if one is confronted with far-IR diagnostics alone,
it is hard to separate shock or turbulently induced gas from a low-density
PDR, since both can exist as a diffuse, warm component in the ISM. 

It is interesting to ask what contribution to the [CII] emission might
be expected from a standard PDR (solar abundance), given the low values of far-UV
field strength G$_o$ of 1.4 Habing units in SQ measured by Guillard et al. (2010). 
Such UV radiation must be present to
ionize the carbon. Based on the models of Kaufman et al. (1999) and
adopting a reasonable density of $\sim$ 10$^3$ cm $^{-3}$, this low
G-field would be expected to contribute a surface brightness in
[CII] emission of 1-2 $\times$ 10$^{-6}$ ergs s$^{-1}$ cm$^{-2}$ sr
$^{-1}$, or between 0.8 and 1.6 $\times$ 10$^{-17}$ W m$^{-2}$, for
the extraction apertures used in our PACS regions. Thus, the UV
radiation field necessary to ionize carbon would contribute
between 10 and 15$\%$ of the emission detected in the SQ filament regions.

Although we will argue below that the metallicity of the gas is not likely to be 
a large factor in explaining the results, we next  consider what the consequences of reduced metallicity might
be in a diffuse PDR. Israel \& Maloney (2011) have detected [CII]/FIR
ratios in some regions of the LMC which range from 1-5$\%$ (Cloud 9 in
LMC-N11 has the highest value), and several regions in the SMC with
elevated values of 0.5-2$\%$, still unusual compared with most
galaxies. These authors argue that PDR models in low metallicity
environments can just, but only just, reconcile the [CII]/FIR ratios
in the LMC. The models of Bakes \& Tielens (1994) can asymptotically
approach values of 5$\%$ for low values of G$_0$. Although one
could argue that in the SQ filament, even lower values of G$_0$ are
likely, the extreme value of [CII]/FIR= 6-7$\%$ seen in the pure-shock
regions of B and E seem to push the model to the limit.  When combined
with the extremely large values of [CII]/PAH discussed earlier, it seems
reasonable to question whether PDR heating of the gas with these large
ratios is the only possible way that [CII] can be heated?

The oxygen metallicity of the gas in the SQ filament has recently
 been measured using optical emission lines with an IFU by
 Iglesias-Paramo et al. (2012).  The results depend on the velocity
 regime being considered. Gas consistent with shock-excitation is
 seen in two main velocity features around 6000-6300 km s$^{-1}$, and
 around 6600 km s$^{-1}$. The low-velocity feature, which is broadly
 associated with the intruder galaxy's (NGC 7318b) velocity (and
 corresponding to the broad left-hand peak seen in both the [CII] and
 CO spectra shown in Figure 2) has close to solar metallicity, but
 the higher velocity gas was found to have an oxygen metallicity of
 12+ logO/H = 8.35, similar to the SMC. However, this result does not
 agree with the earlier measurements of Xu et al. (2003) who found 12
 + log(O/H) = 8.76--slightly super-solar. The differences may be due,
 in part, to the different methods used, but may also be due to the
 difficulty of measuring metallicity when the lines are broad and
 faint. If we accept that the higher velocity component in the [CII]
 and CO profiles could be of lower metallicity, how might this effect
 our interpretation of the global ratios, such as [CII]/PAH and
 [CII]/FIR? The answer is that the effect is minimal, since the
 average metallicity of the gas, integrated over the whole profile,
 is likely to be no less than 0.75 solar, especially as the
 lower-velocity component dominates the integral flux of the [CII] in
 most of the cases shown in Figure 2. As mentioned earlier, no
 large differences in the I([CII])/I(CO J=1-0) ratio between the low
 and high velocity components are obvious, again suggesting that if
 the higher-velocity component in the [CII] profiles has reduced
 metallicity, it must also be compensated for in the CO abundance.

We note that \citet{sarg12} have noticed large
[CII]/PAH ratios, particularly in galaxies with known AGN.
In this case, extinction in the mid-IR was cited
as a possible explanation, although no deep silicate features were
observed. Furthermore, Guillard et al. (in preparation) have discovered similar
cases in radio galaxies which already have been shown to contain
shocked molecular hydrogen. Thus, although rare, galaxies with extreme
[CII]/PAH ratios are beginning to emerge from the {\it Herschel} data. Could shock-heating,
or cosmic rays  accelerated in the shock be the answer? 

\subsection{Cosmic Ray Heating?}

We have shown that the bulk of the gas mass in the shock has to be
warm (T $>$ 50 K) to account for the observed high [CII] cooling
rate. Since
we have ruled out UV, and X-ray heating, we are left with two possible heating
sources of the gas. The first is Cosmic Rays (CR), and the second is the
dissipation of kinetic energy through turbulence and shocks. 

The SQ intergalactic filament was discovered through its synchrotron
radiation in the centimeter waveband \citep{all72,thij81}, and thus
cosmic rays are present in the filament. Following the same calculation
done in \citet{nes10,Og10}, we compute the
CR ionization rate needed to heat the observed gas mass at
T$\sim$100K.  For CRs, the heating energy per ionization is $\sim$
13 eV for H$_2$ gas \citep{gla12}.  To heat the gas we need
an ionization rate per nucleon $\zeta$ of 1 to 2.5 $\times$ 10$^{-14}$ s$^{-1}$.

Each ionization destroys one H$_2$ molecule.  For the gas to be
molecular, the creation rate of H$_2$ ($\gamma_{H_2}$$\times$n(HI)$\times$n(H))
should be higher than the destruction rate $\zeta$$\times$n(H$_2$), where
$\gamma_{H2}$ is the H$_2$ formation rate coefficient.  From Milky Way
observations $\gamma_{H2}$ $\sim$ 1--3 $\times$ 10$^{-17}$ cm$^3$
s$^{-1}$ (Jura 1975).  If the molecular fraction
of the gas is f$_{H2}$ = 2$\times$n(H$_2$)/n(H), then the gas can be molecular if
$\gamma_{H2}$ $>$ $\zeta$ $\times$ f$_{H2}$/(2n(H)$\times$(1-f$_{H2}$)). For the SQ shock,
f$_{H2} >$ 0.95, (based on the lack of HI detected in the filament), and  so
$\gamma_{H2}$ $>$ 9.5$\times$$\zeta$/n(H).  For such a high molecular fraction
we would need n(H) $>$$>$ 1 $\times$ 10$^3$ cm$^3$ for molecules to survive 
if the heating comes from CRs, unless the formation rate of H$_2$
per collision is much larger than the Galactic value. For example, if we assume n(H) is 1000 cm$^{-3}$ 
then $\gamma_{H2}$ would be  1.9 $\times$ 10$^{-16}$ cm$^3$ s$^{-1}$. \footnote{We note that the above argument is very simplistic regarding H$_2$ formation, since, in reality, many complex factors affect the formation rate (e. g. Hollenbach \& McKee 1979, Cazaux \& Tielens 2002, Pirronello et al. 1997), and as we argue later, shocks may accelerate H$_2$ formation through both gas-phase and grain processes.} We therefore conclude that in order for CR heating to be viable, it would require much higher densities of the molecular gas than is observed.

\subsection{Shocks and Turbulence as the likely Heating Source}

We believe that the most likely heating source for the gas in the filament is 
the dissipation of
kinetic energy released through a turbulent cascade
from a large scale (set by the size of the intruder galaxy and its high velocity) to a small scale, with most of the dissipation being at
small scales. This heating source was proposed by Guillard et al. (2009)
to account for the warm H$_2$ emission in SQ, and shown to be viable
for the similar heating of warm H$_2$ in the bridge between the Taffy
galaxies (Peterson et al. 2012).

There is no shortage of available energy to be dissipated in a high
velocity collision between, in this case, NGC 7318b and a pre-existing
HI filament. The turbulent heating rate per unit H$_2$ molecule,
$\Gamma _{turb}$ = (3/2)$\times$ 2m$_H$($\sigma^2$) / $\tau$$_d$,
where $\tau$$_d$ = L / $\sigma$ is the dissipation timescale. Here
$\sigma$ is the characteristic velocity dispersion of the gas, L
is the characteristic length for the system, and m$_H$ is the mass of the hydrogen atom. Given the observed
velocity widths of typically $\Delta$V (FWHM) = 400-800 km/s, then $\sigma$ =
$\Delta$V(FWHM)/2.36 for a Gaussian distribution, lies in the range 170-340 km/s. For a galaxy colliding
with gas clumps with scales of order 4-8 kpcs, the dissipation time is
$\tau$$_d$ $\sim$25 Myrs, and $\Gamma _{turb}$ = 2-8 $\times$
10$^{-31}$ W molecule$^{-1}$, or approximately 4-5 times the cooling
rate of the C$^+$ ions (5-8 $\times$ 10$^{-32}$ W H$^{-1}$--see Fig.6),
and a similar magnitude for the warm H$_2$.

\begin{figure}
\includegraphics[width=8.0cm]{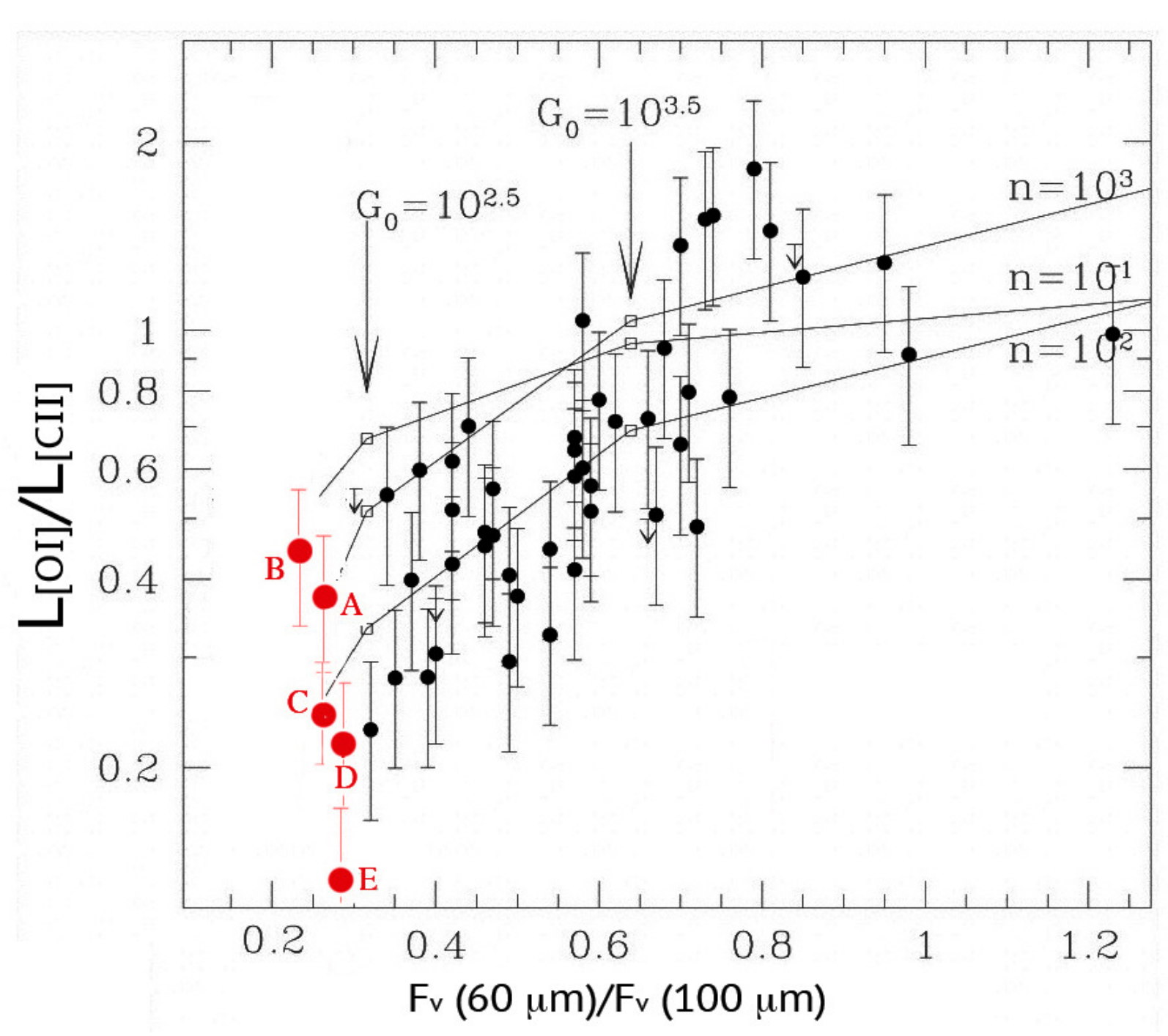}

\caption{\footnotesize{The line luminosity ratio of [OI]/[CII] as a function of the far-IR
color from a sample of galaxies by \citep{malh01} based on ISO
data (black points) with the PACS Regions 1-5 for SQ plotted as red
filled circles.   The solid lines are theoretical models of PDRs
based on the models of  Kauffman et al. (1999). Unlike the other figures
shown in this paper, the SQ points appear to form a continuous
distribution at cool color temperatures with the normal galaxies (see
text).
}}
\end{figure}

Thus turbulence is quite capable of delivering energy at the right rate
for a significant period of time. This energy is likely to be funneled 
to the smallest
scales through a network of low-velocity shocks (Guillard et al 2009;
Lesaffre et al. 2013) or intermittent vortices (Godard et al. 2009) via
supersonic turbulence.

In a remarkably predictive paper, \citet{Les13}, demonstrate that even
quite low velocity shocks passing through a diffuse (10$^2$--10$^3$
cm$^{-3}$) irradiated molecular medium can significantly boost the
[CII] signal so that it becomes almost a bright as the 0-0 H$_2$ S(1)
line.  These authors, who were working with previously
published H$_2$ excitation diagrams for SQ and diffuse gas in the
Chamaeleon cloud, showed that the H$_2$ excitation in SQ could be
modeled by low-velocity C-shocks in a diffuse molecular medium. When
considering mildly UV-irradiated gas (sufficient to ionize the carbon)
they showed that in 7 -10 km/s C-shocks, the [CII] emission can become
almost a bright as the H$_2$ emission. In their
model, 70$\%$ of the [CII] emission comes from shocked
gas with only 30$\%$ from the PDR associated with the ionzation of the carbon. 
Furthermore, these models predict weak [OI]63$\mu$m emission,
largely because of the much lower compression of C-shocks versus more
powerful J-shocks. Although their model underpredicts the ratio of [CII] to
0-0 S(0) H$_2$ line, a network of low-velocity C-shocks is a promising direction for explaining the observations presented here. 

Excluding the SQ-A star formation region, the ratio of I([CII])/I(CO
J=1-0) is observed to lie between 1000 and 1700 in the filament. We
have shown that taken together with the high values of [CII]/FIR
ratio, these values are not consistent with PDR models.

A proper model of the I([CII])/I(CO) ratio for the shock case
is non-trivial, and is very dependent on the specific conditions,
including magnetic field, density and the treatment of the turbulent
line broadening which affects the optical depth, and is beyond the
scope of the present paper. The Lesaffre et al. models somewhat under-predict the I([CII]/I(CO J=1-0) ratios observed here, but were optimized for a different set of observational constraints.

Preliminary shock
modeling, which includes  post-processing with a Large Velocity Gradient (LVG) code, 
is able to reproduce I([CII])/I(CO J=1-0) ratios of 700-3000 for
pre-shock densities of 10$^3$ cm$^{-3}$. 
Future models of the SQ system, incoporating these kinds of models, 
will also have to take into account
energy input to the gas from much higher velocity shocks, which Cluver
et al. (2010) suggested may help to explain the relatively strong
[SiII]34.8$\mu$m line--cooling in the filament, as well as possible
cooling in important far-UV lines.

\section{H$_2$O detection: Evidence for a denser warm molecular component at intermediate radial velocities}

The discovery of H$_2$O from the shock at several positions indicates
that some parts of the shock must contain high densities $>$ 10$^{6}$
cm$^{-3}$, and are likely the result of higher-velocity J-shocks. 
Furthermore, the failure to
detect other lines, on either the otho- or para-water branches is
consistent with a relatively low temperature $\sim$ 80-100 K based on a
simple exploration of parameter space in the RADEX models of Van de
Tak et al. (2007). Such high densities are in contrast to the density needed to explain the [CII] emission, supporting the idea of the SQ
filament being composed of several distinct gas phases.  The line
luminosity in the p-H$_2$O (1$_{11}$-0$_{00}$) line is 10$\%$ of the [CII] line
intensity in region SSWD4 which lies close to PACS Region C, and is
sampled at almost the same scale. We note
that in an unpublished GBT search for 22 GHz H$_2$O masers in the
filament, none were found (P. Appleton, Personal Communication), which
suggests that ultra-dense gas ($>$ 10$^{7-8}$ cm$^{-3}$), needed for
maser action, is rare in the structure, or alternatively, that the observer
is not aligned with the maser column. 

As we have noted, the width of the p-H$_2$O line is unresolved at the
FTS resolution of 324 km s$^{-1}$ and has a heliocentric velocity in
the mid-range of [CII] velocities detected in the filament. At
the position of the SSWD4 footprint (see Fig.1c), the velocity of the line
lies at 6457($\pm$ 90) km s$^{-1}$, which is BETWEEN the velocity of the
intruder galaxy NGC 7318b (V = 5774 km s$^{-1}$) and the gas at the
main velocity of the group (V = 6600-6700 km s$^{-1}$). This places it
at the same velocity as the majority of the warm H$_2$ emission
\citep{m10}, and between the peaks of the CO 1-0 profile seen in the
central panel of Fig.2. Thus the H$_2$O line may be gas that
represents the material accelerated by the intruder galaxy to intermediate
velocities. Perhaps only this strongly accelerated gas is permeated by
J-shocks. 

It is worth discussing a very different interpretation of the line
seen in the FTS spectrum of Fig.3. Could the line be identified,
not with the p-H$_2$O line, but with ortho-H$_2$O$^+$ (1$_{11}$-0$_{00}$)  
which is seen in absorption in M82 (Wei{\ss} et al. 2010)? 
o-H$_2$O$^+$ has a transition at a rest
frequency $\nu$ = 1115.186 GHz (M{\"u}rtz et al. 1998), which would be blue-shifted by 
507 km s$^{-1}$ compared with p-H$_2$O. This would place the line at a
velocity of $\sim$6000 km$^{-1}$, close to the main peak in the CO and
[CII] profile for Region C. A search for other H$_2$O$^+$ lines (for example the one
at 1108.404 and 1132.629) yielded no detection. Nevertheless, could the
detection of o-H$_2$O$^+$ be viable?

As discussed by \citep{hol12}, H$_2$O$^+$ is part of
the ``backbone'' of interstellar chemistry in the ISM, and is a
main formation route from H$_2$ to H$_2$O. In order for oH$_2$O$^+$ to
be detected (without a detection of H$_2$O), the water would have to
be very highly ionized. Models of the abundance of ionized water under
various conditions of photoionization and cosmic ray ionization are
presented in Figs.2 and 3 of Hollenbach et al. What is noticeable is
that in almost all cases considered, the abundance of OH$^+$ is
at least comparable, or larger than H$_2$O$^+$, especially in the more
diffuse environments. These authors consider CR ionization rates of up
to 2 $\times$ 10$^{-16}$ s$^{-1}$, which is comparable to the limiting
ionization cases considered in Section 5.4. At this CR ionization
rate, H$_2$O would also be strong, and yet only one line is detected.
The non-detection of the OH$^+$ lines at 903.684, 966.92 and 1026.903
GHz \citep{bek85} may argue against a highly ionized medium, and the H$_2$O$^+$
interpretation of the 1113 GHz line. 

\section{Implications for Turbulent gas at high-z}

The discovery of kinematically--broad group-wide [CII] emission that
appears too bright to be explained by star formation (either arising
in HII regions or by diffuse PDRs) in
Stephan's Quintet is a strong reminder that [CII] can be relatively
easily enhanced by processes other than UV heating. The fact that we
see the signal without strong contamination from PDR emission, as well
as sharing a similar velocity space with CO (1-0) emission, suggests a
close connection with the dissipation of kinetic energy in the group
as the intruder crashes through the intergroup gas. The spatial extent
of the observed [CII] emission covers most of the observational area of
our PACS observations, and undoubtably extends over the whole of the
shocked filament and the bridge to NGC 7319. The large width of the
[CII] line (in some places $>$ 1000 km s$^{-1}$) as well as its
comparable power to the H$_2$ 0-0S(1) line suggests that a [CII]
emission originating in a highly turbulent medium. A turbulent cascade 
from large-scale high velocity shocks created
in the galaxy collision, to small-scale low-velocity shocks in the H$_2$ gas
can explain a large part of the [CII] emission. 

Ogle et al. (2010) and Guillard et al. (2012) have already
demonstrated that radio galaxies are often hosts of very strong warm
H$_2$ emission, and that shocks and turbulence are the main cause 
(see also Nesvadba et al. 2011 for additional evidence for shocks in 3C326)
Furthermore, very high H$_2$ line luminosities have recently
been found associated with the z = 2 ``Spiderweb'' radio galaxy (Ogle et al. 2012).  Such strong
emission is potentially associated with shocks within 60 kpc of the radio
galaxy and surrounding satellite systems, which already shows strong
filamentary H$\alpha$ emission. Thus objects like the ``Spiderweb''
may be scaled-up versions of the SQ phenomenon, and would be ideal
places to look at higher-z for enhanced [CII] emission. In this respect, it
is interesting that Seymour et al. (2012) claim a possible photometric detection of [CII] in their 500$\mu$m band at a tentative level of [CII]/FIR = 2$\%$. 
Follow--up of this possible discovery with line observations would be very 
interesting.   

Observatories like ALMA are now capable of detecting highly redshifted
[CII] emission from galaxies, sometimes with
higher than expected [CII]/FIR ratios \citep[e. g. ][]{swi12}.
Although one explanation for the enhanced [CII] emission is that
the star formation is quite extended, and the UV-field diluted, our observations
of SQ provide an alternative possibility--that at least some of the 
extra [CII] comes from the dissipation of mechanical energy in the systems. 
As we illustrate this in Fig.9, the regions of SQ which contain known
star formation show lower values of [CII]/FIR.
There is no reason to believe that SQ's filament is
unique--it is simply one of the best known nearby examples of an
isolated extragalactic shocked structure, quite separated from regions of strong
star formation. We are currently reducing data on the Taffy bridge
system, which is also known to contain a strong shock signature
(Peterson et al. 2012), and preliminary indications are that the
bridge also emits strong [CII] emission. It is clear from our energy dissipation discussion in Section 5.5, that turbulence can deliver large amounts of energy to gas on a timescale comparable with star formation, and should not be neglected in violently collisional situations. In an environment where
dark-matter halos are scaled up (e. g. in massive radio galaxies or
quasars) more energy will be available for both star formation and
dissipation of mechanical energy. 

Unfortunately it is not clear how to separate the UV-heated gas from
the shock-heated gas without more diagnostics than those available from
a single high-z detection of [CII]. Even if [OI] is also detected, as
we showed in Fig.10, it is not easy to separate diffuse PDR emission
from shock-induced [CII] emission, because both can lead to a diffuse, low
density medium. In addition, there is the possibility of
AGN heating of the [CII], which, like higher velocity shocks, will
also boost the [OI] emission. The detection of strong molecular
hydrogen cooling lines in high-z galaxies, with the proposed {\it
  SPICA} telescope, or ro-vibrational lines of H$_2$ with {\it JWST}, may
provide additional hope for separating PDR heating from
turbulence and shocks.

\section{Conclusions}

We have performed the first {\it Herschel} observations of the giant
intergalactic filament away from major regions of star formation in
the compact group, Stephan's Quintet. Previous observations have
provided strong evidence that the filament is the result of the
supersonic collision between an intruder galaxy, NGC7318b and a
pre-existing tidal tail in the group. Our observations with the PACS
and SPIRE spectrometers lead to the following conclusions:

\begin{itemize}

\item{Kinematically broad ($>$1000 km s$^{-1}$) far-IR fine structure
lines of [CII]$\lambda$157.74$\mu$m are detected from the giant
shocked filament and bridge in Stephan's Quintet. Weaker
[OI]$\lambda$63.18$\mu$m emission is also detected. The [CII]
emission is very extensive, and can be decomposed into two or
three different velocity components, many of which have FWHM
ranging from a few hundred km s$^{-1}$ to 733 km s$^{-1}$. 
The [CII] emission profiles at various positions in the 
filament are similar to CO 1-0 emission detected with IRAM in
similar areas, suggesting that the molecular gas and the [CII]
emission are closely related.}

\item{Based on a model of the heating and cooling of the [CII] and [OI]-emitting gas, our measurements of the cooling rates in the filament suggest the emission arises in a warm diffuse medium (90 $<$ T $<$ 200 K, 10$^3$ $<$ $\rho$ $<$ 3 $\times$ 10$^3$ cm$^{-3}$). The temperature and mass of the gas involved is consistent with the main collisional partner of the C$^+$ ion being the same warm molecular gas previously detected by {\it Spitzer} observations of the rotational H$_2$ lines, which is believed to be shock-heated.

\item{The [CII] line emission exceeds by a factor of two in most cases
the strength from the 0-0 S(1) warm molecular hydrogen line, and
is typically 50-60$\%$ of the total rotational line
luminosity from molecular hydrogen (0-0S(0) to 0-0S(5)). In some regions [OI]63$\mu$m carries about the same power as the 0-0S(1) line, whereas
p-H$_2$O is 28$\%$ of the 0-0S(1) line in the area where the SPIRE
and PACS detectors overlap. Thus the far-IR line-cooling from the 
giant filament is significant. Warm molecular hydrogen is still the 
dominant coolant over the 6 rotational lines detected by {\it Spitzer.}}

\item{Based on a careful comparison with previous {\it Spitzer} spectral 
mapping in the mid-IR, we find extremely large values of the ratio 
[CII]/PAH$_{7-8}$, ranging from 0.3 - 3, and [CII]/PAH$_{total}$, ranging 
from 0.17 - 0.7. Similarly, unusually large value of [CII]/FIR ranging 
from 0.04 to 0.07 were found in the same regions using imaging data from 
{\it Herschel}  and {\it Spitzer} to constrain the SED. The [CII]/PAH 
ratios are an important measure of the heating efficiency of the diffuse 
ISM by photo-electric heating, and the observed values are so large that 
it is unlikely that most of the observed [CII] is emitted from PDRs.
Furthermore, we can rule 
out direct UV excitation from HII regions and X-ray heating. 
The strong enhancement of [CII] in 
the filament, relative to PAH and dust continuum, suggests strong heating of 
the [CII] from shocks and turbulence, and not star formation.}

\item{Emission is also detected in several places from the p-H$_2$O (1$_{11}$0$_{00}$) line including the center of the giant
filament. Unlike the [CII] and [OI] emission from that position,
the H$_2$O emission is unresolved ($\Delta$V $<$ 360 km s$^{-1}$),
and has a systemic velocity placing it in the mid-range of
detected [CII] emission and at the same velocity are strong warm
H$_2$ detected by {\it Spitzer}. The detection of H$_2$O suggests 
that the SQ filament contains regions of very high gas density 
($>$ 10$^6$ cm$^{-3}$), supporting the idea that the structure of the filament is highly multi-phase.  
Despite detection of the CO 1-0,
2-1 and 3-2 lines from the ground, no higher-order J lines of CO
or other H$_2$O lines were detected with SPIRE.}  

\item{We considered
the possibility that the detected SPIRE line was H$_2$O$^+$, not
H$_2$O, but rejected this possibility because the high
ionization rates needed to completely ionize water, would also
be expected to ionize OH, forming OH$^+$, which is not detected.}

\item{Models \citep{Les13} of mildly (G$_o$$\sim$1) UV-irradiated
molecular gas show that the [CII] signal can be strongly boosted without
a larger increase in [OI] emission in low-velocity magnetic shocks. Although these
models do not explain all the emission detected, the work strongly suggests shocks are playing a large role. In reality, the SQ filament is likely
permeated by a vast network of shocks (as evidenced by the broad line-widths),
including numerous low-velocity
magnetic (C-) shocks traveling through molecular gas of relatively
low density (100-1000 at cm$^{-3}$)--a density which falls within the range 
we derive from the observations. These low-velocity shocks could be the natural consequence of turbulent energy dissipation initiated on much larger scales by the galaxy collision in the group.} 
   
\item{The possibility that [CII] can be excited on a large scale by
the dissipation of mechanical energy (turbulence and shocks)
provides a potential new source of [CII] emission when
interpreting [CII] in high-redshift galaxies
\citep[e. g. ][]{gord10,swi12}.  Since, in this paper, there is
strong evidence for a connection between the turbulent molecular
gas in the SQ shocked filament, and the [CII] emission (both its
distribution and kinematics), we suggest that enhanced [CII] (and
warm H$_2$ emission) from turbulent energy dissipation should be
present in most situations where highly turbulent conditions
exist--such as in galaxy collisions, and in the early stages of the
building of galaxy disks. However, as we have shown, the energy
appears to be degraded down to very small scales and
low-velocities, creating a warm diffuse gas component which is not
easily distinguished from a low-density PDR.  Thus, the best
places to disentangle mechanically-induced [CII] emission 
from UV-dominated heating will be in
regions where turbulent processes are present, but star formation
has not yet turned on. Such conditions may exist in the early
stages of galaxy formation.}
}
\end{itemize}

PNA would like to acknowledge interesting discussions with
P. Goldsmith and W. Langer (JPL) regarding [CII] emission in the Galaxy. 
This work is based on observations made with {\it Herschel}, a European Space Agency Cornerstone Mission with significant participation by NASA. Support for this work was provided by NASA through an award issued by JPL/Caltech. The authors wish to thank an anonymous referee for thoughtful comments on a previous version of the text.

\clearpage

\begin{table}
{\scriptsize \tiny

\caption{Line Fluxes of Selected Regions (See  1b and c) }

\begin{center}
\begin{tabular}{c l l l l l l l l l}
\hline
\hline
\\[0.5pt]
PACS & RA & DEC & S(1)H$_2$ & H$_2tot$ & PAH$_{(7-8)} $ & PAH$_{tot}$ & H$_2tot$/PAH$_{7-8}$ & H$_2tot$/PAH$_{tot}$ & H$_2tot$/FIR$^e$ \\
Region$^{a}$  & (J2000)   & (J2000)    & Flux$^{b}$  & Flux$^{b,c}$ & Flux$^{b,d}$ & Flux$^{b,e}$ & & & \\
\hline
\\
A & 22 35 58.9 & 33 58 52.8 & 3.96 (0.07) & 11.44 (0.29) & 32.4 (4.4) & 59.5 (7.5) & 0.35 (0.05)  & 0.19 (0.02) & 0.060 (0.004) \\    
B & 22 35 58.9 & 33 58 29.9 & 4.39 (0.07) & 14.55 (0.28) & 6.2 (1.3) & 17.38 (4.2) & 2.3 (0.5) & 0.8 (0.2) & 0.132 (0.014)\\ 
C & 22 35 59.8 & 33 58 05.3 & 4.44 (0.07) & 17.36 (0.29) & 13.3 (3.3) & 25.2 (6.1) & 1.3 (0.3) & 0.69 (0.16) & 0.124 (0.012) \\
D & 22 35 59.8 & 33 57 40.9 & 2.91 (0.07) & 10.54 (0.29) & 18.0 (3.3) & 31.3 (6.6) & 0.59 (0.11)  & 0.34 (0.07) & 0.082 (0.010) \\
E & 22 36 01.6 & 33 58 25.9 & 3.86 (0.07) & 12.19 (0.29) & 2.34 (1.1) & 10.44 (3.7)  & 5.2 (2.5) & 1.2 (0.4) & 0.106 (0.013) \\
\hline
\\
PACS & RA & DEC & [CII] Line & [OI]63$\mu$m & [CII]/H$_{2tot}$ & [CII]/PAH(7.7) & [CII]/PAH$_{tot}$ & [CII]/[OI] & [CII]/FIR$^{f}$  \\
Region$^{a}$ & (J2000) & (J2000) &  Flux$^{b}$ & Flux$^{b}$ & ratio$^{c}$ & ratio$^{d}$ & ratio$^{e}$  & & \\
A & 22 35 58.9 & 33 58 52.8  & 9.65 (1.5)  & 3.9 (0.7)   & 0.84 (0.13)   & 0.29 (0.06)  & 0.16 (0.03) & 2.5 (0.77) & 0.047 (0.008)  \\
B & 22 35 58.9 & 33 58 29.9  &  9.02  (1.4) & 4.1 (1.0)  & 0.62  (0.09)   & 1.45 (0.37)  & 0.52 (0.15) & 2.22 (0.64) & 0.078 (0.013) \\ 
C & 22 35 59.8 & 33 58 05.3  &  10.84 (1.6) & 2.5 (0.6)   & 0.62 (0.09)  & 0.81 (0.24) & 0.43 (0.12) & 4.34 (1.27) & 0.074 (0.012) \\
D & 22 35 59.8 & 33 57 40.9  &  5.59  (0.8) & 1.2 (0.5)$^g$     & 0.53 (0.08)   &  0.31 (0.07) & 0.18 (0.05) & 4.6 (2.0) & 0.041 (0.007) \\
E & 22 36 01.6 & 33 58 25.9  &  7.34  (1.1) & 0.8 (0.4)$^g$     & 0.60 (0.09)  &  3.1 (1.5) & 0.71 (0.27)  & 9.2 (4.8) & 0.061 (0.010) \\

\hline
\\
SPIRE & RA& DEC & p-H$_2$O & S/N & Peak/rms & & & & \\  
Region$^{a}$ & (J2000) & (J2000) & 111-000  & & & & & & \\
             &         &         & Flux$^{b}$ & & & & & & \\
SSWD4 & 22 36 00.4  & 33 58 04.2 & 1.24 (0.13) & 9.6 & 4.9   & detected & & & \\
SSWE3 & 22 36 01.2  & 33 58 36.8 & 1.25 (0.19) & 6.7 & 3.4 & detected & & & \\ 
SSWE4 & 22 35 58.7  & 33 58 29.7 & 1.44 (0.19) & 7.9 & 4.0 & detected & & & \\
SSWC4 & 22 35 59.6  & 33 57 32.4 & 1.06 (0.15) & 7.0 & 3.6 & detected & & & \\
SSWC3 & 22 36 02.0  & 33 57 40.5 & 0.73 (0.16) & 4.5 & 2.3 & Marginal detection & & & \\ 
\hline
\\[0.5pt]
\multicolumn{8}{l}{$^a$ Regions defined in Fig. 1: each covering 18.8 x 18.8 arcsecs$^2$} \\
\multicolumn{8}{l}{$^b$ Fluxes in units of 10$^{-17}$ W m$^{-2}$} \\
\multicolumn{8}{l}{$^c$ Sum over the line luminosities 0-0S(0) to 0-0S(5)}\\
\multicolumn{8}{l}{$^d$ Sum over 7.3-8.7$\mu$m PAH complexes} \\
\multicolumn{8}{l}{$^e$ Sum over 6.3-17$\mu$m PAH bands}\\
\multicolumn{8}{l}{$^f$ FIR fluxes based on integral over SED  (see Table 4 and text).}\\
\multicolumn{8}{l}{$^g$ Marginal detections of [OI].} \\

\\[0.5pt]
\\
\end{tabular}
\end{center}
}
\end{table}

\begin{table}
{\scriptsize \tiny
\caption{Velocity Components of the Detected [CII], [OI] and H$_2$O Lines }
\begin{center}
\begin{tabular}{c l l l l l l l l l l l l}
\hline
\hline
\\[0.5pt]
Name & Line & Ncomps & V$_1$ & $\delta$V$_1$ & Fluxfrac$_1$ & V$_2$ & $\delta$V$_2$ & Fluxfrac$_2$ & V$_3$ & $\delta$V$_3$ & Fluxfrac$_3$ & Comment \\
     &      &        & km s$^{-1}$  & km s$^{-1}$ & (percent) & km s$^{-1}$ & km s$^{-1}$ & (percent) & km s$^{-1}$ & km s$^{-1}$ & (percent) & \\
\hline
\\
A & [CII] & 2 & 6093 (120) & 591 & 30.4 & 6748 (120) & 378  & 69.6 & -- & --  & -- & Poor fit  \\    
A & [CII]  & 3 & 6101 (80)& 383 & 32 & 6726 (80) & 319  & 60.0 & 7024 (80) & 234 & 8 & Good \\
A & [OI] 63$\mu$m   & 2 & 6097 (100) & 342 & 35   & 6719 (100) & 512  & 65 & -- & -- & -- & Fair      \\
B & [CII] & 3 & 6121 (80) & 415 & 56 & 6490 (80) & 333 & 22  & 6829 (80) & 437 & 22 & Good \\ 
C & [CII] & 2 & 6041 (80) & 475 & 47 & 6539 (80) & 770 & 53 & -- & -- & -- & Good \\
D & [CII] & 2 & 5819 (120) & 357 & 50  & 6377 (120) & 637 & 50 & -- & -- & -- & Poor \\
D & [CII] & 3 & 5781 (80) & 281 & 34  & 6107 (80)  & 540 & 45  & 6592 (80) & 331 & 21 & Good \\
E & [CII]  & 2 & 6078 (80) & 234 & 10 & 6355 (80)  & 640  & 90  & -- & -- & -- & Good  \\
SSWD4 & p-H$_2$O (1$_{11}$0$_{00}$) & 1 & 6457 (120) & $<$360   & 100 & -- & -- & -- & -- & -- & -- & Sinc function \\
SSWE3 & p-H$_2$O (1$_{11}$0$_{00}$) & 1 & 6425 (120) & $<$360   & 100 & -- & -- & -- & -- & -- & -- & Sinc function \\
SSWE4 & p-H$_2$O (1$_{11}$0$_{00}$) & 1 & 6481 (120) & $<$360   & 100 & -- & -- & -- & -- & -- & -- & Sinc function \\
SSWC4 & p-H$_2$O (1$_{11}$0$_{00}$) & 1 & 6380 (120) & $<$360   & 100 & -- & -- & -- & -- & -- & -- & Sinc function \\
SSWC3 & p-H$_2$O (1$_{11}$0$_{00}$) & 1 & 6353 (120) & $<$360   & 100 & -- & -- & -- & -- & -- & -- & Marg. Detection \\
\hline
\\
\\[0.5pt]
\\
\end{tabular}
\end{center}
}
\end{table}

\begin{table}
{\scriptsize \tiny

\caption{Upper Limits to Important Undetected Lines}

\begin{center}
\begin{tabular}{c l l l l l l l}
\hline
\\
SPIRE & LINE & Rest $\nu$ & RA & DEC & Line Flux$^{b}$ & Beam FWHM & Comment \\  
DETECTOR & &(GHz) &  (J2000) & (J2000) & & (arcsecs) & \\

SLWC3 & CO 4-3 & 461.040 & 22 36 00  & 33 58 04  & $<$1.1 & 46 & 3-sigma upper limit  \\
SLWC3 & CO 5-4 & 576.267 & 22 36 00  & 33 58 04  & $<$0.46 & 37 & 3-sigma upper limit \\ 
SLWC3 & CO 6-5 & 691.473 & 22 35 00  & 33 58 04  & $<$0.27 & 31 & 3-sigma upper limit  \\
SLWC3 & CO 7-6 & 806.651 & 22 35 00  & 33 58 04  & $<$0.21 & 26 & 3-sigma upper limit  \\
SLWC3 & CO 8-7 & 921.800 & 22 35 00  & 33 58 04  & $<$0.34 & 23 &3-sigma upper limit  \\
SLWC3 & pH$_2$O 2$_{02}$1$_{11}$ & 987.927 & 22 35 00 & 33 58 04 & $<$0.48 & 21 & 3-sig upper limit \\
SLWC3 & pH$_2$O 2$_{11}$2$_{02}$ & 752.033 & 22 35 00 & 33 58 04 & $<$0.26 & 28 & 3-sig upper limit \\
SLWC3 & oH$_2$O1$_{10}$1$_{01}$   & 556.936 & 22 35 00 & 33 58 04 & $<$0.47 & 38 & 3-sig upper limit \\    
SSWD4 & [NII]205$\mu$m & 1461.132 & 22 35 00.4 & 33 58 04.2 & $<$0.27 & 14  & 3-sigma upper limit \\
SSWD4 & OH$_c$$^+$ & 1033.004$^c$ & 22 35 00.4  & 33 58 04.2 &  $<$0.45 & 21 & 3-sigma upper limit \\
SLWC3 & OH$_a$$^+$ & 909.045$^c$  & 22 35 00.4  & 33 58 04.2 & $<$0.30 & 23 & 3-sigma upper limit \\
SLWC3 & OH$_b$$^+$ & 971.803$^c$  & 22 35 00.4  & 33 58 04.2 & $<$0.28 & 22 & 3-sigma upper limit \\

\hline
\\[0.5pt]
\multicolumn{8}{l}{$^a$ Regions defined in Fig. 1} \\
\multicolumn{8}{l}{$^b$ Fluxes in units of 10$^{-17}$ W m$^{-2}$} \\
\multicolumn{8}{l}{$^c$ OH$^+$ rest frequencies from \citet{bek85}. } \\

\\[0.5pt]
\\
\end{tabular}
\end{center}
}
\end{table}

\begin{table}
{\scriptsize \tiny
\caption{Far-IR SED Fitted Properties for PACS Regions$^a$}
\begin{center}
\begin{tabular}{c l l l l l l l} 
\hline
\hline
\\[0.5pt]
\hline
PACS & $\alpha$$^{b}$ & T$_{dust}$ & FIR Flux$^{c,e}$ & Log(L$_{FIR}$)$^f$ & $\nu$F$_{70}$/$\nu$F$_{100}$ & LOG(M$_{dust}$) & LOG(M$_{gas}$  \\
REGION$^d$ & & (K) & (42-122$\mu$m)$^{g}$ & (10$^{35}$W) & & (M$_{\odot}$) & (M$_{\odot}$)  \\       
\hline
\\
A  & 1.9 (0.1) & 22.9 (0.5) & 1.92 (0.13) & 2.02 (0.13) & 0.55 (0.05) & 6.02 (0.06) & 8.24 (0.09)  \\
B  & 2.1 (0.1) & 22.7 (0.7) & 1.10 (0.12) & 1.16 (0.13) & 0.52 (0.05) & 5.82 (0.08) & 8.04 (0.10)  \\
C  & 2.1 (0.1) & 23.7 (0.6) & 1.40 (0.14) & 1.47 (0.14) & 0.53 (0.05) & 5.83 (0.07) & 8.05 (0.01)  \\
D  & 2.0 (0.1) & 24.0 (0.8) & 1.29 (0.15) & 1.36 (0.15) & 0.55 (0.05) & 5.76 (0.08) & 8.05 (0.11)  \\
E  & 1.5 (0.1) & 21.8 ( 0.8) & 1.15 (0.14) & 1.21 (0.14) & 0.63 (0.06) & 5.82 (0.08) & 8.05 (0.01)   \\
\hline
\\
\\[0.5pt]
\multicolumn{8}{l}{$^a$Using the SED fitting method of Casey(2012).} \\
\multicolumn{8}{l}{$^{b}$ Parameters from Casey(2012). Dust properties assume a dust emissivity of the form $\nu$$^{1.8}$, and power law index $\alpha$.} \\ 
\multicolumn{7}{l}{$^c$derived from {\it Herschel} photometry observations described in companion paper--Guillard et al. (2013).} \\
\multicolumn{7}{l}{$^d$PACS Regions as defined in Fig. 1b and Table 1 covering 18.8 x 18.8 arcsecs$^2$.} \\
\multicolumn{7}{l}{$^e$Fluxes in units of 10$^{-15}$ W m$^{-2}$.} \\
\multicolumn{7}{l}{$^f$Assuming D = 94 Mpc.} \\
\multicolumn{7}{l}{$^g$See Dale \& Helou 2003.} \\
\\
\\[0.5pt]
\\
\end{tabular}
\end{center}
}
\end{table}

\begin{table}
{\scriptsize \tiny
\caption{H$_2$ Properties derived from Spitzer IRS Observation for each PACS Region$^a$}
\begin{center}
\begin{tabular}{c l l l l l l l l}
\hline
\hline
\\[0.5pt]
PACS &  N1(H$_2$) & T1 H$_2$ & OPR$^b$ & M1$_{H2}$ & N2(H$_2$) & T2 H$_2$ & OPR &  M2$_{H2}$ \\
REGION$^a$ & x (10$^{19}$ cm$^{-2}$) & K & &  M$_{\odot}$ (x 10$^7)$ & x (10$^{17}$ cm$^{-2}$) & K & &  M$_{\odot}$ (x 10$^5)$   \\       
     &  & & & & &  \\
\hline
\\
A &  14.0 & 157 & 2.6 & 15.2 & 6.6 & 628 & 3 & 7.1    \\
B &  5.9 & 187 &  2.8 & 6.3  & 7.4 & 666 & 3 & 8.0   \\
C &  4.0 & 201 &  2.9 & 4.3  & 8.8 & 676 & 3 &9.5   \\
D &  2.0 & 217 &  2.9 & 2.2  & 6.9 & 617 & 3 & 7.4   \\
E &  6.7 & 175 &  2.7 &7.2  & 5.9 & 663 & 3 & 6.3   \\
\hline
\\[0.5pt]
\multicolumn{7}{l}{$^a$ PACS Regions as defined in Fig. 1b and Table 1 each covering 18.8 x 18.8 arcsecs$^2$.} \\
\multicolumn{7}{l}{$^b$ Ortho to Para ratio determined from the fit to the H$_2$ excitation diagrams for each position.} \\ 

\\
\\[0.5pt]
\\
\end{tabular}
\end{center}
}
\end{table}

\begin{table}
{\scriptsize \tiny
\caption{X-ray Properties of PACS Regions}
\begin{center}
\begin{tabular}{c l l l l l l l l l} 
\hline
\hline
\\[0.5pt]
\hline
PACS & kT$^a$ & Z$^a$ & L$_{Xsoft}$$^{a,c}$ &  L$_{Xsoft}$/L$_{FIR}$) & [CII]/L$_{Xsoft}$ & H$_{2tot}$/L$_{Xsoft}$\\
REGION$^b$ & (keV) & (solar) & (10$^{33}$W) & (10$^{33}$W)  & & \\       
\hline
\\
A  &  0.39 ($^{+0.13}_{-0.06}$) & 1.33 ($^{+1.33}_{-1.20}$) &  0.90 ($^{+0.02}_{-0.02}$) &  0.005 & 11.1 (1.7) & 13.4 (0.4) \\
B  &  0.62 ($^{+0.06}_{-0.05}$) &  0.12 ($^{+0.06}_{-0.04}$) &  3.31 ($^{+0.25}_{-0.25}$) & 0.029 & 2.9 (0.5) & 4.6 (0.4) \\
C  &  0.71 ($^{+0.05}_{-0.09}$) &  0.18 ($^{+0.09}_{-0.06}$) & 3.35 ($^{+0.25}_{-0.25}$) &  0.023 & 3.4 (0.6) & 5.5 (0.4) \\
D  &  0.70 ($^{+0.07}_{-0.09}$) &  0.45 ($^{+2.04}_{-0.26}$)  & 1.66 ($^{+0.18}_{-0.18}$) &  0.012 & 3.6 (0.6) & 6.7 (0.8) \\
E  &  0.60 ($^{+0.07}_{-0.08}$) & 0.15 ($^{+0.12}_{-0.06}$) &  2.19 ($^{+0.21}_{-0.21}$) &  0.018 & 3.5 (0.6) &  5.9 (0.6)  \\
\hline
\\[0.5pt]
\multicolumn{7}{l}{$^a$Soft X-rays (0.5-2keV) derived from extractions from the  {\it CHANDRA} observations by O'Sullivan et al. (2009).} \\
\multicolumn{7}{l}{$^b$PACS Regions as defined in Fig. 1b and Table 1 covering 18.8 x 18.8 arcsecs$^2$.} \\
\multicolumn{7}{l}{$^c$Assuming D = 94 Mpc.} \\
\\
\\[0.5pt]
\\
\end{tabular}
\end{center}
}
\end{table}

\end{document}